\documentclass[10pt,a4paper,twocolumn]{article}
\usepackage[utf8]{inputenc}
\usepackage{amsmath}
\usepackage{amsfonts}
\usepackage{amssymb}
\usepackage{titlesec}
\usepackage{graphicx}
\usepackage{fullpage}
\usepackage{authblk}
\usepackage{booktabs}
\usepackage[backend=biber,style=numeric,sorting=none]{biblatex}
\usepackage{subcaption}
\usepackage{rotating}
\usepackage{hyperref}
\usepackage{cleveref}

\bibliography{main}

\title{Equivariant Graph Neural Networks for Prediction of Tensor Material Properties of Crystals}

\author{Alexander J. Heilman$^*$}
\author{Claire Schlesinger$^*$}
\author{Qimin Yan}
\affil{Northeastern University}
\date{}

\renewenvironment{abstract}
 {\small
  \begin{center}
  \bfseries \abstractname\vspace{-.5em}\vspace{0pt}
  \end{center}
  \list{}{
    \setlength{\leftmargin}{.5cm}%
    \setlength{\rightmargin}{\leftmargin}%
  }%
  \item\relax}
 {\endlist}

\begin{document}
\maketitle

\def\thefootnote{*}
\footnotetext{These authors contributed equally to this work}
\def\thefootnote{\arabic{footnote}}

\begin{abstract}
 Modern E(3)-Equivariant networks may be used to predict rotationally equivariant properties, including tensorial quantities. Three such quantities: the dielectric, piezoelectric, and elasticity tensors, are computationally expensive to produce ab initio for crystalline systems; however, with greater availability of such data in large material property databases, we now have a sufficient target space to begin training equivariant models in the prediction of such properties. Here we explicitly develop spherical harmonic decompositions of these tensorial properties using their general symmetries. We then apply three distinct E(3)-equivariant convolutional structures to the prediction of the components of these decompositions, allowing us to predict the aforementioned tensorial quantities in an equivariant manner and compare performance. We further report results testing the transferability of these predictive models between different tensorial target sets.
\end{abstract}
\vspace{.5cm}

\section{Introduction}
Traditional  machine learning methods applied to the material sciences have predicted invariant, scalar properties of material systems to great effect \cite{mlreview1, mlreview1.5, mlreview2, mlreview2.25, mlreview2.5, mlreview3}. Equivariant models now may predict coordinate system dependent outputs in a well defined manner \cite{tensorfieldnetworks, e3nn_paper}, though some applications neglect the prediction of directional (i.e. coordinate system dependent) quantities and instead are used to predict invariant quantities to greater accuracy \cite{alignn}.

In rotationally equivariant models, features are often associated with irreducible representations of the rotation group \cite{e3nn_paper}, generally referred to as spherical harmonics. This association may be leveraged to predict tensorial quantities directly from the outputs of such models.
This component-wise prediction of tensorial properties is achieved by decomposing tensors into harmonic subspaces via a \textit{spherical harmonic decomposition}, by which we may also associate arbitrary tensors with the irreducible representations of the rotation group \cite{harmonictensors}. This allows one to read off tensors component-wise from the output representations of these equivariant models by matching rotational and azimuthal indices.

In this work, we present results for the prediction of various material property tensors directly from crystalline structures. Namely, given some material's crystalline structure, we predict tensor components of dielectric, piezoelectric, and elasticity tensors directly from the output of an $SE(3)$ equivariant model.



This is similar to the approach taken in \cite{yan2024equivcrystalprediction}, where equivariant outputs were also used to predict tensorial quantities. Other works, such as StrainNet \cite{StrainNet} have previously used equivariant graph transformers and additional information about the strained structure of a crystal to predict elasticity tensors. MatTen \cite{MatTen} utilized a steerable equivariant convolutional neural network to predict the elastic tensor with a decomposition based on the same principles demonstrated here. 

Below, we give explicit decompositions of these material property tensors (with a brief overview of the mathematics involved in attached appendices), generally following the approach taken in \cite{itin-elastic, itin-rank3}. We then present results for some basic implementations of this approach, and further attempt some transfer learning applications.

\section{$SO(3)$ Equivariant Neural Networks}
Neural networks are a class of universal function approximators, composed layer-wise by functions $\mathcal{L}^1\circ\mathcal{L}^2\circ ... \circ \mathcal{L}^n $, where each layer-to-layer transition map $\mathcal{L}^i$ is a trainable function from some feature space associated with layer $L$ to a feature space associated with layer $L+1$.

Some approaches use a specific type of neural network, referred to as a Graph Neural Network (GNN) \cite{cgcnn,alignn,amdnet,chemgnn,geocgcnn,icgcnn} or more generally a message passing network (MPNN) \cite{mpnn}, which acts on data encoded in features associated with some representative graph (i.e. a collection of nodes and connections between them). These models preserve the underlying connectivity of the graph representation, and act in a way that is invariant under node-index permutation.
Basic graph networks are also often rotationally invariant, however. As such, these models are incapable of predicting any coordinate-system quantities, such as vector tensor components. 

Equivariant functions are those that preserve the action of a group on it's domain and codomain. For a function $f:X\rightarrow Y$, where $X$ and $Y$ are vector spaces with some set of group representations $\mathcal{D}^{X}(G)$ and $\mathcal{D}^Y(G)$ acting on these spaces respectively, $f$ is equivariant with respect to $G$ if it satisfies:
$$
\mathcal{D}^{Y}(g)f(x)=f\big( \mathcal{D}^{X}(g)x\big)
$$
Compositions of equivariant functions are themselves equivariant functions  \cite{equivariant_cohen}. As such, we may form an equivariant network by composing it layer-wise from a set of common equivariant functions.

In material systems, we generally desire representations that are equivariant under the group of transformations acting on real space. This corresponds to the special Euclidean group $SE(3)$, which consists of translations of the origin and the group of proper rotations $SO(3)$ \cite{e3nn_paper}. MPNNs are generally invariant under translations of the origin based upon the features of the graph representations. As long as only interatomic distance is encoded and atomic positions are treated accordingly, if included. 

In an $SO(3)$-equivariant network, features are further associated with $SO(3)$ irreducible representations: namely, spherical harmonics $Y_{\ell}^m$, which are doubly indexed by rotational order $\ell\geq 0$, and azimuthal order $-\ell\leq m \leq \ell$.
In such cases, a traditional feature set $V^{(n)a}$ of channel $a$ and associated with object $n$, has an additional two indices $\ell$ and $m$ in an $SO(3)$ network \cite{tensorfieldnetworks}, corresponding to the aformentioned indices of the harmonic representations.

\begin{figure}[t]
\centering
\includegraphics[width=\linewidth]{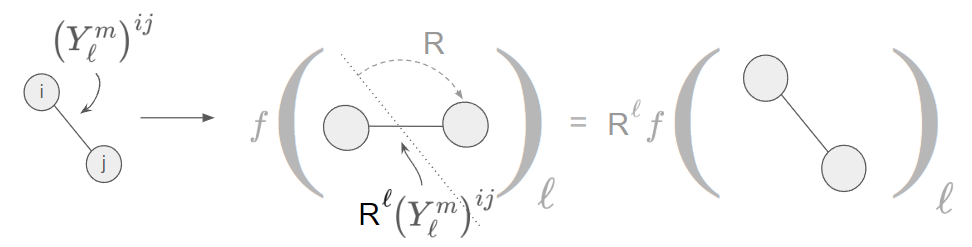}
\caption{Example diagram showcasing a rotationally equivariant update scheme where features are associated with spherical harmonics (indexed by $\ell$ and $m$). Note that the function $f$'s output also has a corresponding $\ell$ index, as well as the rotation operator $R^{\ell}$.}
\end{figure}

Here, we consider four types of $SO(3)$-equivariant functions from which we may compose our equivariant networks layer-wise, following \cite{equivariant_cohen}. These consist of: $SO(3)$-feature convolutions, $\ell$-wise self-interactions and non-linearities, and pooling across nodes. A brief overview of each is given below.

\subsection*{$SO(3)$-Equivariant Convolution}
Features and filters in $SO(3)$ networks are both associated with $SO(3)$ representations, or spherical harmonics.
Tensor products of these two representation spaces are equivariant under transformation of the two subspaces, i.e.:
$$
\mathcal{D}^V\otimes \mathcal{D}^W=\mathcal{D}^{V\otimes W}.
$$
By way of Clebsch-Gordan coefficients $c^{\ell_3m_3}_{\ell_1m_1\ell_2m_2}$, these tensor products of $SO(3)$ irreducible representations may be related to a third set of irreducible representations as \cite{sakurai}:
$$
(u\otimes v)_{\ell_o}^{m_o} = c_{\ell_1m_1\ell_2m_2}^{\ell_om_o}u_{\ell_1}^{m_1}v_{\ell_2}^{m_2}
$$
where $u$ and $v$ are harmonic vectors of order $\ell_1$ and $\ell_2$, respectively.

Since tensor products of representations are naturally equivariant, we may define $SO(3)$-equivariant convolution to be the scaled tensor product of the two representation spaces (i.e. that of the input feature space, and the filter space). Specifically, layer to layer convolutional maps $\mathcal{L}$ may be defined component-wise as\cite{tensorfieldnetworks}:
$$
\mathcal{L}^{\ell_o}_{acm_o}\big(\vec{r}_a,V_{acm_i}^{\ell_i}\big) = \sum_{m_f,m_i}c_{\ell_im_i\ell_fm_f}^{\ell_o m_o}\sum_{b}F^{\ell_f\ell_i}_{cm_f}(r_{ab})V_{bcm_i}^{\ell_i}
$$
where the filter function $F^{\ell_f\ell_i}_{cm_f}(r_{ab})$ depends only on the distance between point $a$ and $b$ (as opposed to directional dependence, to maintain equivariance). Instances of such functions then generally have independent, trainable parameters for different rotational orders $\ell_f, \ell_i$, azimuthal orders $m$, and channels $c$.


\subsection*{Self-Interaction}
Feature sets for individual objects may also update according to themselves as long as they act across $m$ for every $\ell$ and only update according to the different channels $c$. That is, functions of the form:
$$
V_{acm}^{\ell} \rightarrow  \sum _{c}W^{\ell}_{c'c}V_{acm}^{\ell}
$$
are also equivariant. Such functions we refer to as \textit{self-interaction} layers.

\subsection*{Non-Linearities}
We can also apply point-wise non linearities and maintain equivariance, as long as they also respect the $\ell$ and $m$ indices of every feature.

\subsection*{Pooling}
Pooling, or aggregation, across all elements or objects (index $a$) while preserving the $m$ and $\ell$ indices is also equivariant. So functions of the form:
$$
 M_{cm}^{\ell} = \text{AGG}_{a}(\lbrace V_{acm}^{\ell}\rbrace)
$$ 
where $\text{AGG}$ is an arbitrary aggregation function performed only over the object index $a$, are also available in the construction of $SO(3)$ networks.

\subsubsection*{SO(3) Equivariant Outputs}
Arbitrary combinations of the above sets of functions may be composed into equivariant models.
The outputs of these networks are then associated with some indices $\ell$ and $m$, which may be interpreted as a set of spherical harmonic basis components.
These $SO(3)$ equivariant networks are naturally well suited for the prediction of tensorial properties. As will be shown below, tensors may generally be decomposed into a set of $SO(3)$ invariant subspaces, which can each be associated with a set of spherical harmonic tensors themselves. Since the outputs of $SO(3)$ networks naturally transform like spherical harmonic tensor coefficients indexed by $\ell$ and $m$, we may then essentially read them off as such and convert these into Cartesian tensor components.

\section{Spherical Harmonic Tensors and $SO(3)$ Invariant Subspaces}\label{so3inv}
We may construct a set of spherical harmonic tensors $\hat{Y}_{\ell}^m$ by first defining the $J_z$ vector basis and then considering symmetric products of such vectors. We define the $J_z$ basis \cite{mochizuki1988spherical}:
$$
\begin{bmatrix}
a_+ \\
a_0 \\
a_-
\end{bmatrix}=\begin{bmatrix}
-\frac{1}{\sqrt{2}} & -\frac{i}{\sqrt{2}} & 0\\
 0 & 0 & 1\\
-\frac{1}{\sqrt{2}} & +\frac{i}{\sqrt{2}} & 0\\
\end{bmatrix}\begin{bmatrix}
x \\
y \\
z
\end{bmatrix}
$$
where $x,y,z$ are Cartesian components of the same vector. In this basis, the unit vectors associated with these components correspond to unit vector spherical harmonics $\hat{Y}_{\ell=1}^m$ (where we will assume Racah normalization for all definitions).

Now, recall the Clebsch-Gordon expansion of products of spherical harmonics (again, in the Racah normalization\footnote{In this work, we define the spherical harmonic functions as: $Y_{\ell}^m(\theta, \phi)=\sqrt{\frac{(\ell-m)!}{(\ell+m!)}}P_{\ell}^m(\cos\theta)e^{im\phi}$; which makes Clebsch-Gordon relations simpler}):
$$
Y_{\ell_1}^{m_1}\otimes Y_{\ell_2}^{m_2} = \sum_{L=-|\ell_1-\ell_2|}^{\ell_1+\ell_2}\sum_{M=-L}^{L}c_{\ell_1 0 \ell_2 0}^{L0}c_{\ell_1 m_1 \ell_2 m_2}^{LM}Y_{L}^M.
$$
This may be used to draw a correspondence between symmetric tensor products of $n$ vectors in the $J_z$ basis and rank $\leq n$ spherical harmonic tensors. These give us two sets of basis tensors for symmetric tensor spaces. The coefficients for such a transformation relevant for the tensors considered here are listed in \cref{cgcoef}.

The spherical harmonics $Y_{\ell}^m$ canonically refer to a complex basis, as should be clear from our definition of the $J_z$ basis. Many tensors describing macroscopic responses of materials are strictly real-valued however. In the applications here we then desire to transform into a set of real-valued spherical harmonics $Y_{\ell m}$, with the transformation adopted here given in \cref{realsph}.

We may decompose an arbitrary tensor of rank-$n$ into a set of symmetric tensors of rank-$\ell$ with $0\leq\ell\leq n$ by projecting it onto it's irreducible $SO(3)$-invariant subspaces. We then may use the relations between the $J_z$ basis components and the real spherical harmonic components $y_{\ell m}$ using the relations discussed above.

$SO(3)$ invariant subspaces can be constructed via a compound decomposition with respect to $GL$, $SL$, and $O$: which result from application of Young symmetrizers, contractions with the totally antisymmetric $\epsilon$, and contractions with the totally symmetric $g$, respectively. For an overview of these decompositions, see \cref{youngoverview}.

\begin{figure*}[t]\label{rank2example}
\begin{center}

\includegraphics[scale=1]{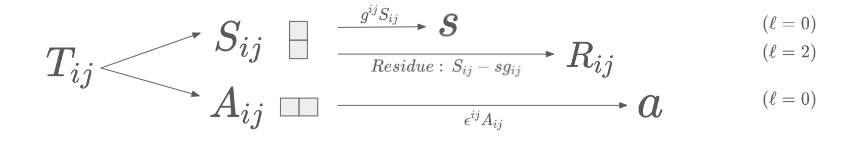}
\end{center}
\caption{Overview of $SO(3)$ decomposition of rank two tensors. Here, $S_{ij}=\frac{1}{2}[T_{ij}+T{ji}]$ is the totally symmetric part of the tensor $T$, and $A_{ij}=\frac{1}{2}[T_{ij}-T{ji}]$ is the totally antisymmetric part. These result from the application of the symmetrizer $\mathcal{S}(ij)$ and the antisymmetrizer $\mathcal{A}(ij)$ acting on $T$, respectively.}
\end{figure*}

As a basic example, consider a general rank-two tensor, as in \cref{rank2example}, where we may first construct the totally symmetric and antisymmetric subspaces, and then contract with the metric tensor $g$ and the Levi-Civita tensor $\epsilon$, respectively. This achieves an irreducible $SO(3)$ decomposition of the space into a trace $s$, a symmetric rank-two residue $R$, and a scalar antisymmetric component $a$.


The relevance of these facts is that, in the manner described above, we are able to decompose an arbitrary tensor into a set of components that transform like spherical harmonics, just like the representations passed throughout, and output by, $SO(3)$ equivariant models. To this end, we now give spherical harmonic decompositions of several tensors describing common material properties, and then use these decompositions to predict these tensors directly from material structure.

\section{Spherical Harmonic Decomposition of Common Material Property Tensors}

 Here is presented a brief overview of the harmonic decomposition of three common, tensorial material properties: the dielectric tensor $\epsilon$, the piezoelectric strain tensor $d$, and the elasticity tensor $C$. A more systematic presentation of each decomposition is given in \cref{decomps}.

\subsection*{Dielectric Tensors}
The dielectric permittivity tensor  $\mathbf{\epsilon}$ of some material is a linear model of it's electric displacement $\vec{D}$ in response to an external electric field $\vec{E}$:
$$
\vec{D}=\mathbf{\epsilon}\vec{E}
$$
Note that here we focus only on static responses, that is, $(\partial/\partial t)\vec{E}=0$, due to the restriction of data to this case. 

The dielectric tensor $\epsilon$ then is a rank-two tensor formed from vector spaces over the field $\mathbb{R}^3$ (where we henceforth ignore the distinction between co- and contravariant components due to the Euclidean structure $g_{ij}=\delta_{ij}$).

The dielectric tensor  $\epsilon$ is symmetric under permutation of it's indices, such that:
$$
\epsilon_{ij}=\epsilon_{ji}.
$$
The dielectric tensor then may be decomposed further into it's trace $s = g_{ij}\epsilon_{ij}$ and it's traceless residue $R$, defined by:
$$
R_{ij}=\epsilon_{ij} - sg_{ij}
$$
resulting in two invariant subspaces of rank-0 and 2, respectively. These two symmetric subspaces then admit a decomposition in the manner described in \cref{so3inv}.

\subsection*{Piezoelectric Tensors}
The piezoelectric strain constants $(d_{ijk})_{T}$ are defined (at constant temperature) by the thermodynamic relation:
$$
\big(d_{ijk}\big)_{T}=\left(\frac{\partial \epsilon_{ij}}{\partial E_k} \right)_{\sigma, T}
$$
where here $\epsilon$ is the strain tensor, and the partial derivative is taken at constant stress $\sigma$ and temperature $T$.

These strain constants $d_{ijk}$ are related to the piezoelectric stress constants $e_{ijk}$ via the elastic tensor $C_{ijkl}$ according to:
$$
\big(e_{ijk}\big)_T=\big(d_{ilm}\big)\big(C_{ijkl}\big)_{E,T}
$$
\begin{center}
with $e_{ijk}$ defined as:
\end{center}
$$
\big(e_{ijk}\big)_{T}=\left(\frac{\partial D_{i}}{\partial \epsilon_{jk}} \right)_{\sigma, T}
$$
and where $D$ is the resulting electric displacement vector in the material (we now generally neglect the explicit notation of constant parameters).

The piezoelectric strain components $d_{ijk}$ are symmetric under $i,j$ due to the symmetry of the strain tensor $\epsilon_{ij}$, so that we have:
$$
d_{ijk}=d_{jik}.
$$
This symmetry gives us a natural $GL$ decomposition into the totally symmetric $S$ and the mixed-symmetry $A$, defined component-wise from $d$ as \cite{itin-rank3}:

\begin{align*}
S_{ijk}&=\frac{1}{3}\big(d_{ijk}+d_{ikj}+d_{kji}\big)\\
A_{ijk}&=\frac{1}{3}\big(2d_{ijk}-d_{ikj}-d_{kji}\big).\\
\end{align*}

The fully symmetric part $S$ then consists of a trace vector $s_{k}=g_{ij}S_{ijk}$, and a symmetric residue $W_{ijk}$ defined by:
$$
W_{ijk} = S_{ijk} - \frac{1}{5}(g_{ij}v_k+g_{ik}v_j+g_{jk}v_i)
$$
These correspond to the spaces  $\mathcal{H}^{(1)}$ and $ \mathcal{H}^{(3)}$, respectively.

The mixed symmetry part $A$ also requires decomposition with respect to $SO(3)$, so that we have a set of symmetric tensors describing it. It's 8 independent components can be described by a $5\oplus 3$ dimensional space consisting of a symmetric rank-2 tensor and a trace vector.

The trace vector $a^i$, describing $A$'s 3-dimensional $SO(3)$ invariant subspace, can be formed from the contraction of the metric tensor $g_{ij}$ along $A$'s first and second indices. That is, we define:
$$
a^i =g_{jk}A^{ijk}
$$
Note that this choice is somewhat arbitrary, since we could define the trace part to correspond to the contraction along the first and third indices, or the second and third. However, it can be shown that for the mixed symmetry of $A$, these two trace vectors are linearly dependent (related by an overall factor of $1$ and $-2$, respectively).

The rank-2 invariant subspace of $A$ then may be constructed by symmetrizing the partial contraction with $\epsilon$ along the first and third indices (the anti-symmetric pair). Note that the antisymmetric part of this partial contraction corresponds to the trace vector space accounted for here by $u$. Explicitly, we define:
$$
b_{ij}=\frac{1}{2}\big(\epsilon_{nim}A_{njm}+\epsilon_{njm}A_{nim}\big)
$$
which is a traceless symmetric rank-2 tensor.

And so, as mentioned above, the symmetric part $S$ is readily decomposed in the spherical bases into $\mathcal{H}^{(1)}\oplus \mathcal{H}^{(3)}$. And then the mixed-symmetry part inhabits the space $\mathcal{H}^{(1)}\oplus \mathcal{H}^{(2)}$.

\subsection*{Elasticity Tensor}

The elasticity (or simply, elastic) tensor $C$ of a material relates it's Cauchy strain $\epsilon$ to some infinitesimal stress $\tau$ \cite{landau-elastic} . This may be described, component-wise, as the linear relation below:
$$
\tau_{ij}=C_{ijkl}\epsilon_{kl}
$$
As should be clear by the number of indices, the elastic tensor is a fourth rank tensor. 

The elastic tensor has several symmetries, but is not totally symmetric in general. For all material systems, $C$ must satisfy the so-called 'minor symmetries' below:
$$
C_{ijkl}=C_{jikl}
$$
$$
C_{ijkl}=C_{ijlk}
$$
resulting from the symmetry of the strain and stress tensors ($\tau_{ij}=\tau_{ji}$ and $\epsilon_{ij}=\epsilon_{ji}$) under the assumption of equilibrium. This reduces the number of independent components from 81 to 36.

For conservative systems in which the elastic deformation is describable in terms of some potential energy function (and which we shall henceforth assume for our application), $C$ has the additional 'major symmetry':
$$
C_{ijkl}=C_{klij}
$$
This further reduces the number of independent components from 36 (from the minor symmetries) to 21 in total.

In a $GL$-decomposition, these symmetries leave us with two non-vanishing subspaces: the totally symmetric $S$, and a mixed symmetry $A$ (\cref{elastic_decomp_app}), defined component-wise as \cite{itin-elastic}:
$$
S_{ijkl}=\frac{1}{3}\big( C_{ijkl} + C_{ikjl} + C_{klij} \big)
$$
$$
A_{ijkl} = \frac{1}{3}\big( 2C_{ijkl} -C_{ikjl} -C_{klij}  \big)
$$
(this mixed-symmetry subspace $A$ also corresponds to Backus' \cite{backus1970geometrical} asymmetric tensor $A$).
The fully symmetric part can of course be converted to spherical harmonic components via a Clebsch-Gordon expansion. The mixed symmetry $A$, however requires further decomposition. 

All six independent components of $A$ can be described by the symmetric (but not traceless) tensor $t$, defined as the double partial contraction of $A$ with the totally antisymmetric tensor $\epsilon$ as:
$$
t_{ij} = \epsilon_{i}^{mk}\epsilon_{j}^{nl}A_{mnkl}
$$
This tensor $t$ then has a harmonic decomposition according to the rank-two Clebsch-Gordon transformation between the $J_z$ basis and the harmonic basis $y_{\ell}^m$. 

As such, the elastic tensor $C$ admits a decomposition into $S$ and $A$, which inhabit spaces of $\mathcal{H}^{(4)}\oplus \mathcal{H}^{(2)}\oplus \mathcal{H}^{(0)}$ and $\mathcal{H}^{(2)}\oplus \mathcal{H}^{(0)}$, respectively.

\section{Methods}

Results presented here investigate the performance of three distinct $SO(3)$ equivariant models in the direct prediction of tensor components from crystal structure. This is achieved in the manner outlined above, wherein we utilize equivariant model's output as a set of spherical harmonic coefficients and then convert these to Cartesian tensor components.

\subsection*{Model Architecture}

\begin{figure}[!t]
\begin{center}
\includegraphics[scale=.47]{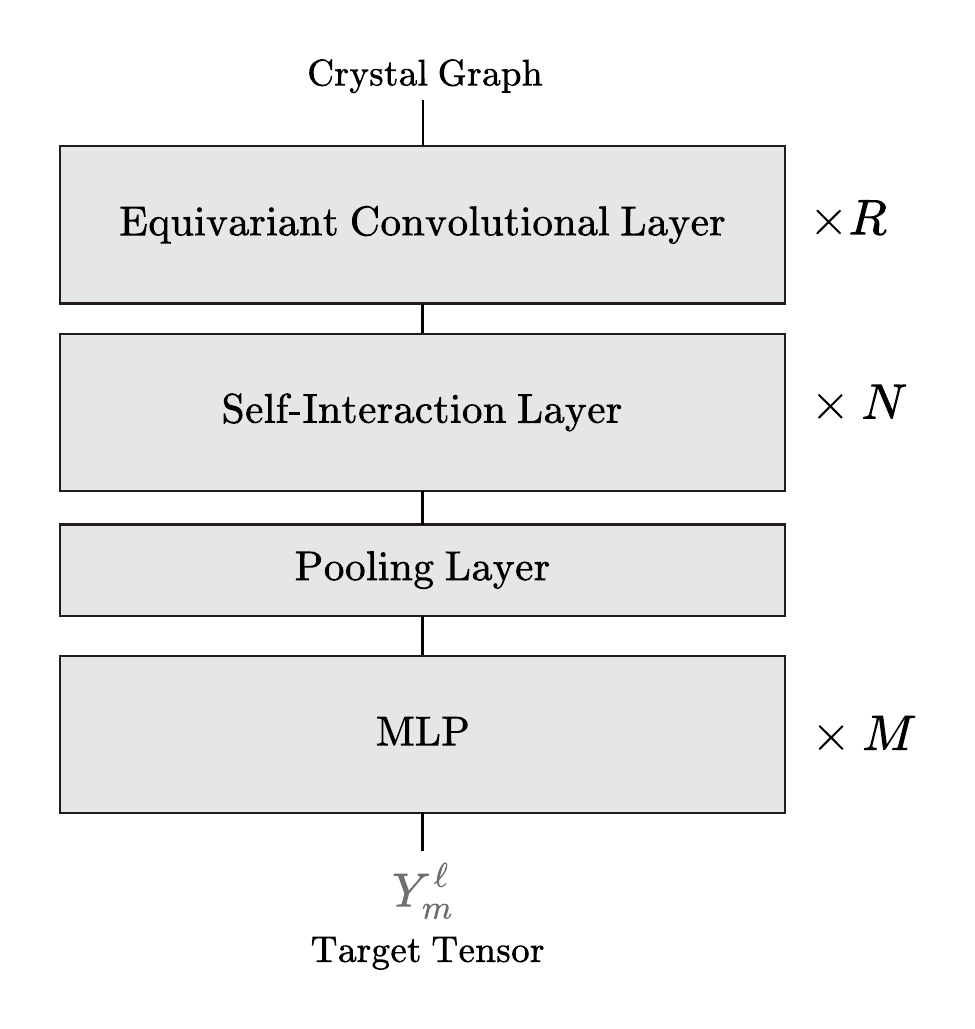}
\end{center}
\caption{Diagram depicting the general architecture of the model used in this work.}
\label{modelarch}
\end{figure}

The model architecture used in this work is based on the design of the SE(3) Equivariant graph neural networks found in \textcite{brandstetter2021geometric}.
This architecture consists of an layer embedding the input atom features into spherical harmonics up to order $\ell_{max}$. These features are then passed through $R$ graph convolution layers (either message passing, convolution, or transformer layers) where nodes are updated by their neighbors. The output representation is then passed through $N$ self interaction layers in which individual nodes update themselves. A pooling layer (either mean or max pooling) is then applied across node indices to return a crystal-level feature vector, which is then passed through $M$ MLP layers. This output, still associated with $\ell$ and $m$ indices, is then used to predict the coefficients of target tensors. A schematic for this architecture is provided in \cref{modelarch}.

Three equivariant convolutional layers were tested and compared in this model architecture: the Steerable Equivariant Graph Neural Network (SEGNN) and the Steerable Equivariant Non-Linear Convolutional layer (SEConv) \cite{brandstetter2021geometric}; and the Steerable Equivariant Transformer (SETransformer) \cite{fuchs2020setransformer}. All are based on the popular e3nn python module \cite{e3nn_paper}, with features and outputs generally associated with a set of spherical harmonic indices. Note that the SETransformer is unique in that it alone uses attention in the filter function $\mathcal{F}_{acm_im_o}^{\ell_o\ell_i}(r_{ab}, v_a,v_b)$, such that edge weights may be learned from origin and connected node features $v$.

\subsection*{Data}

The data utilized is available in the \textit{Materials Project} database \cite{MPMain} which includes $\sim$150k materials. Of which, almost all have calculated scalar values for band gap and formation energy, $\sim$12k have a calculated elastic tensor, $\sim$6k have a calculated  dielectric tensor, and $\sim$3k have a calculated piezoelectric tensor. The data was split  70\% for training and 15\% for validation, with a remaining 15\% held out for testing. Data distribution statistics can be found in \cref{data-dist}.

Graphs input into the model were derived from the material structures available in the database. For up to the first twelve neighbors of atoms within 8 Angstrom, edges were constructed. These edges' features consisted of spherical harmonics of orders $0\leq\ell\leq \ell_{max}$ evaluated on each edge's corresponding unit vector direction, with the distance between atoms being included as $\ell=0$ features. Initial node features then encoded the following atomic information: atomic number\footnote{\label{onehot}These properties were one-hot encoded.}, group\footref{onehot}, period\footref{onehot}, block\footref{onehot}, electronegativity, atomic radius, number of valence electrons, electron affinity, ionization energy, and atomic mass; following the method adopted in \cite{cgcnn}.

\section{Results}
Here we present results for the prediction of tensor components from a model of the form discussed above, and compare the relative performance of three distinct equivariant convolutional layers in \cref{pred}.

One significant hurdle in the prediction of tensorial material properties is the scarcity of training data. Pretraining on a diverse set of related targets may be particularly useful in this case \cite{pretraining}. To test the transferability of tensor data to different tasks, we train convolutional layers through a variety of target set progressions. Models are first pretrained on one of two scalar datasets: band gap or formation energy, and then compared to un-pretrained benchmarks in \cref{trans-pred}. These are then compared to cascades of pretraining on different tensorial target sets in \cref{chain-pred}. Finally, a set of common scalar elastic values are calculated from the predicted elasticity tensors, with such predictions discussed in \cref{scalar-pred}.

\subsection{Tensor Prediction}\label{pred}
\begin{figure}[!t]\label{best-heatmap}
\begin{center}

\includegraphics[width=\linewidth]{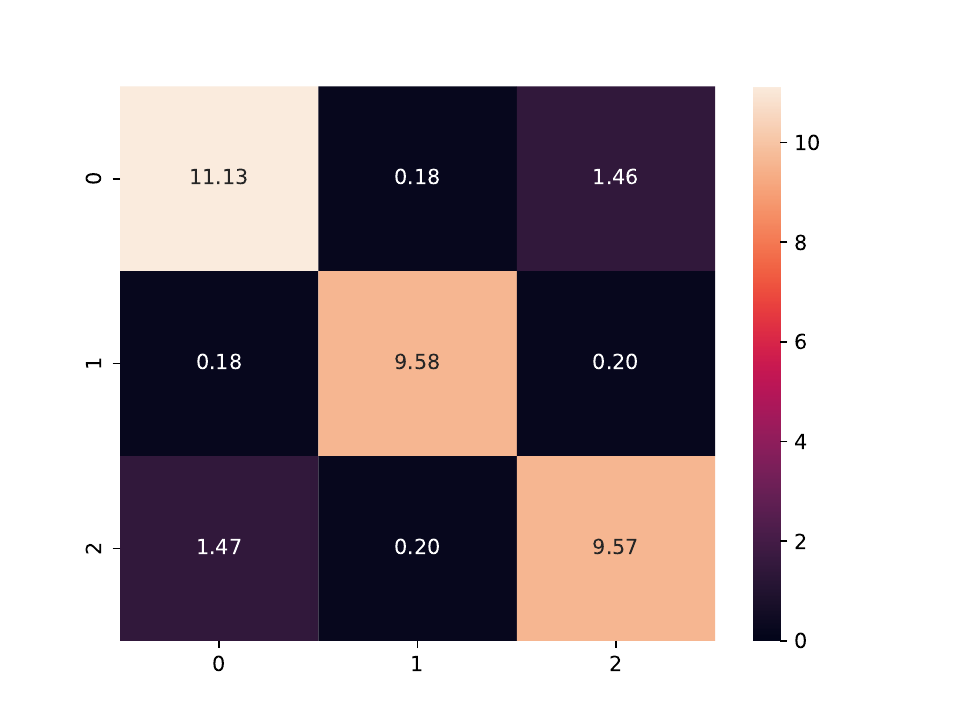}
\includegraphics[width=\linewidth]{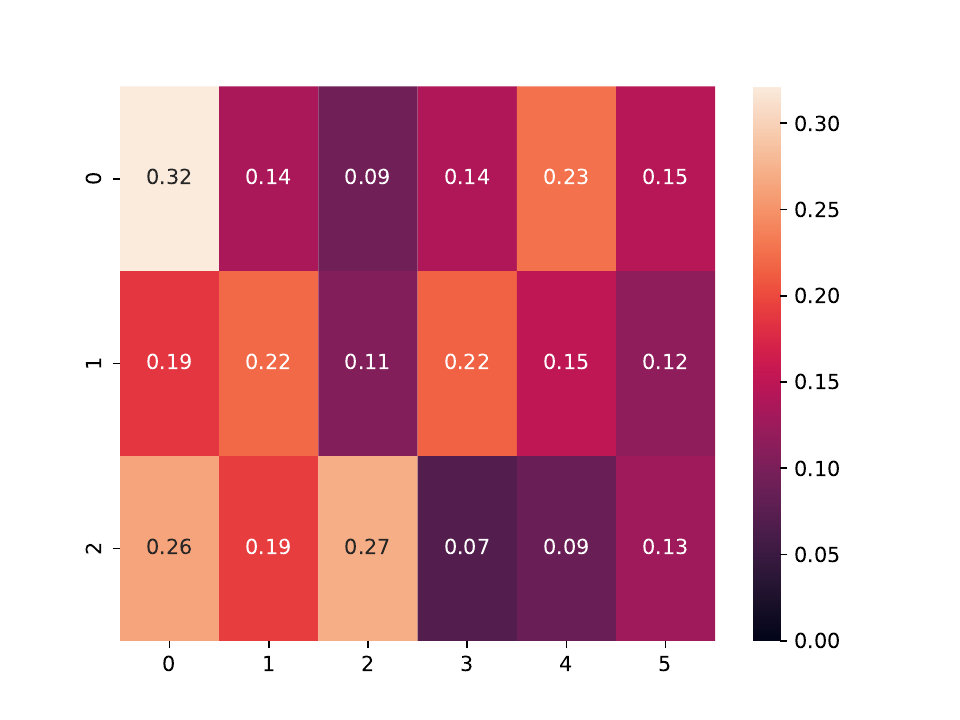}
\includegraphics[width=\linewidth]{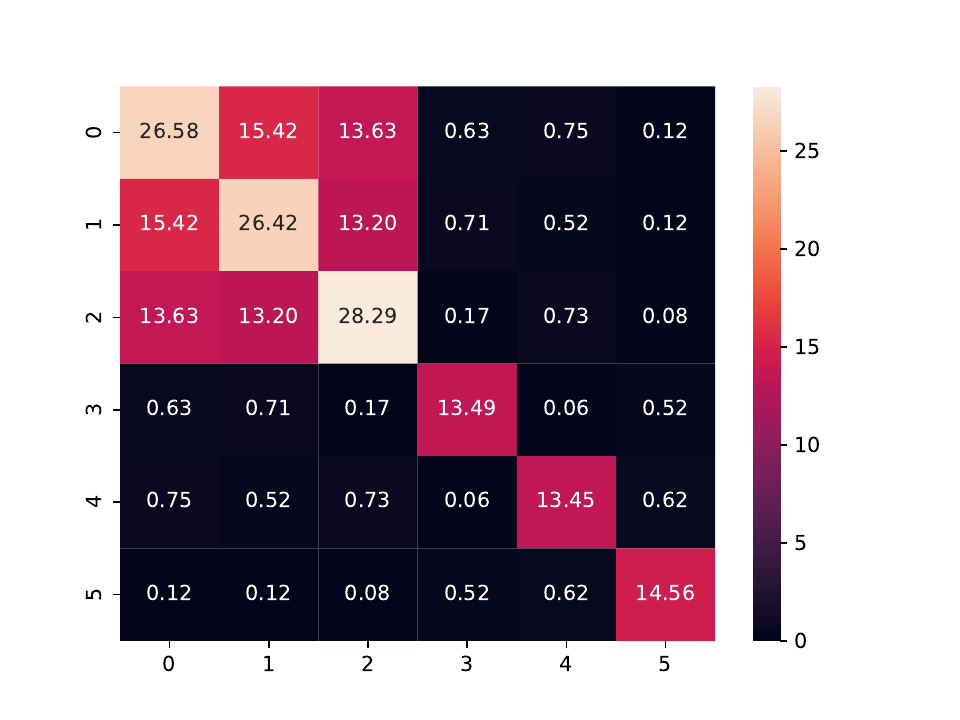}
\end{center}
\caption{Heatmaps showing the component-wise MAE in Voigt notation for the all direct prediction tasks. Shown in order of dielectric, piezoelectric, and elastic tensor components, respectively. These prediction statistics are from a model with SEGNN convolutional layers.}\label{heatmaps}
\end{figure}

Three distinct models were trained to predict tensor targets component-wise from a set of randomly initialized parameters, and then tested on a set of test data withheld through training. Test results for these predictions are given in \cref{direct}, where metric used to report results for prediction tasks is average MAE over components, following the choice of metric for such predictions in \cite{StrainNet}. Component-wise MAE is presented in \cref{heatmaps}.

The average MAE across tensor components in testing beats previous results as reported in \cite{StrainNet}, with an improvement from $10.37 \ GPa$ to $7.69 \ GPa$ with SEConv models in this work. Best results for other tensorial targets were $4.7$ with the SEConv model, and $0.17\ C/m^2$ with both SEGNN and SEConv. 

Since some components must be zero for high-symmetry systems, these components may be unfairly influential in a naive scalar metric based on an average across all components for all systems. Indeed, heatmaps displaying component-wise MAE in \cref{best-heatmap} suggests incorporation of symmetry class information is necessary to further improve results, as in \cite{yan2024equivcrystalprediction}. This should be apparent in the lower MAE contributions in several off-diagonal components in the dielectric and elasticity prediction heatmaps, where in an average over all systems, many must be zero due to symmetry constraints.

\begin{table*}[t]
\centering
\begin{tabular}{|c|ccc|}
\hline
& \multicolumn{3}{|c|}{MAE (Averaged Over Components)} \\
Target Components & SEGNN & SEConv & SETransformer\\
\hline 
Elastic  \textit{[GPa]}& 8.139 & 7.689 & 7.941 \\
Dielectric & 4.82 &4.702 &4.718\\
Piezoelectric \textit{[C/$m^2$]} & 0.170 & 0.170 & 0.1714\\
\hline
\end{tabular}
\caption{Results for direct prediction of elastic, piezoelectric, and dielectric tensor components.}\label{direct}
\end{table*}

\subsection{Scalar Pretraining \& Transfer Learning Applications}\label{trans-pred}
Transfer-learning experiments here focus on a filter-down approach, in which we start training on larger datasets and then transfer the trained graph network weights to a downstream task with a smaller dataset (while continuing to train the transferred network).

Since all tensorial data is relatively scarce, pretraining on a scalar target set would be ideal, since such data is the most plentiful for material systems. To test such applications, we here consider training progressions of the following form:
$$
\textit{Scalar Target} \rightarrow \textit{Tensor Target},
$$
The two scalar datasets used in pretraining included the predicted formation energy per atom for a crystal, and the calculated band gap.

Results from these experiments can be found in \cref{scalar2tensor}. Note that in these progressions, no data was witheld in the scalar target pretraining process. In general, pretraining of this nature had little effect on down-stream results. For predictions after pretraining on the scalar formation energy, the prediction accuracy across components only improved substantially for the SETransformer model: with average MAE decreasing to $7.65\ GPa$ (from $7.94\ GPa$) for elasticity components, and $4.41\ $ (from $4.72$) for dielectric targets. Pretraining on band gap had even less effect on down-stream performance, except for a notable improvement in the SEConv model's dielectric average test MAE from $4.7$ to $4.35$.  Other predictions had little-to-no improvement after scalar pretraining.

Poor-to-no improvement after such pretraining  may be due to a number of factors: perhaps there is little similarity between pretraining and end target datasets, so that pretraining overgeneralizes the model to a larger class of materials than are relevant. Alternatively, the difference in the rotational order of outputs between tasks may also result in little overlap between the pretrained filter weights and the weights relative to the down-stream task. This suggests pretraining on different tensor targets of similar order may improve down-stream task performance more than scalar pretraining sets. 

Poor transfer-learning results may also be due to little domain relevance between different targets. This is a point reinforced by the most substantial improvements being observed in arguably more related target set pairs: namely, from formation energy to elasticity predictions in the SETransformer; and the pretraining progression of band gap to dielectric predictions in the SEConv progression.

\begin{table*}[t]
\centering
\begin{tabular}{|c|c|ccc|}
\hline
&  & \multicolumn{3}{|c|}{MAE after pretraining (\textit{without pretraining})} \\
\textit{Pretraining Task:} & \textit{Tensor Target}  & SEGNN & SEConv & SETransformer\\
\hline
 & Elastic \textit{[GPa]} & 8.19 \textit{(8.14)} &  7.98 \textit{(7.69)} & 7.65 \textit{(7.94)}\\
Formation Energy & Dielectric & 5.02 \textit{(4.82)} & 4.52 \textit{(4.70)} & 4.41 \textit{(4.72)} \\
& Piezoelectric \textit{[C$/m^2$]} & 0.170 \textit{(0.170)} & 0.170 \textit{(0.170)} & 0.171 \textit{(0.171)}\\
\hline
 & Elastic \textit{[GPa]} & 8.39 \textit{(8.14)} &  7.87 \textit{(7.69)} & 8.34 \textit{(7.94)}\\
Band Gap & Dielectric & 4.74 \textit{(4.82)} & 4.35 \textit{(4.70)} & 4.56 \textit{(4.72)} \\
& Piezoelectric \textit{[C$/m^2$]} & 0.170 \textit{(0.170)} & 0.170 \textit{(0.170)} & 0.171 \textit{(0.171)}\\
\hline
\end{tabular}
\caption{Results for tensor component prediction accuracy  after pretraining on a scalar-target task for  materials. Numbers italicized in parentheses are the original accuracies (without pretraining).}\label{scalar2tensor}
\end{table*}

\subsection{Chained Pretraining for Piezoelectric Tensors}\label{chain-pred}
The smallest dataset considered here is that of piezoelectric strain constants. Performance did not significantly improve with any scalar target pretraining in any of the models for this task. The cost and scarcity of data for this target motivates further search for improvement from other tensorial target sets.

To this end, the experiments performed here focus on three different training progressions of tasks:
\begin{enumerate}
\item Formation energy/Band gap $\rightarrow$ Elastic $\rightarrow$ Dielectric $\rightarrow$ Piezoelectric
\item Formation energy/Band gap $\rightarrow$ Elastic  $\rightarrow$ Piezoelectric
\item Formation energy/Band gap $\rightarrow$ Dielectric $\rightarrow$ Piezoelectric
\end{enumerate}
The performance of these chained pretraining progressions in the final prediction of piezoelectric components is reported in \cref{experimentpiezo}. 

Overall, there was no change in average prediction accuracy for the piezoelectric tensor through all experiments of the above form, with an average MAE remaining $0.17\ C/m^2$ for all. The reasons for this lack of improvement may be similar to those given in the scalar pretraining discussion: disjointed target data, and/or filter weights due to differing rotational orders between tasks. However, there should be atleast some overlap in filter weight training due to the common prediction of order $\ell=2$ outputs between all three tensor targets. It may be that the scarcity of piezoelectric target data alone is the fundamental restriction on performance, in which case a larger set of down-stream training data would improve the effectiveness of such pretraining strategies.

\begin{table*}[t]
\centering
\begin{tabular}{|c|c|ccc|}
\hline
&  & \multicolumn{3}{|c|}{MAE (Avg. Over Components) \textit{[C$/m^2$]}} \\
\textit{Scalar Dataset:}& \textit{Experiment \#}& SEGNN & SEConv & SETransformer\\
\hline
& Exp. 1 & 0.170 \textit{(0.170)} & 0.170 \textit{(0.170)} & 0.171 \textit{(0.171)}\\
Formation Energy & Exp. 2 & 0.170 \textit{(0.170)} & 0.170 \textit{(0.170)} & 0.171 \textit{(0.171)}\\
& Exp. 3 & 0.170 \textit{(0.170)} & 0.170 \textit{(0.170)} & 0.171 \textit{(0.171)}\\
\hline
& Exp. 1 & 0.170 \textit{(0.170)} & 0.170 \textit{(0.170)} & 0.171 \textit{(0.171)}\\
Band Gap & Exp. 2 & 0.170 \textit{(0.170)} & 0.170 \textit{(0.170)} & 0.171 \textit{(0.171)}\\
& Exp. 3 & 0.170 \textit{(0.170)} & 0.170 \textit{(0.170)} & 0.171 \textit{(0.171)}\\
\hline
\end{tabular}
\caption{Results for prediction accuracy on the piezoelectric tensor after chained pretraining.}\label{experimentpiezo}
\end{table*}

\begin{table*}[t]
\centering
\begin{tabular}{|c|ccc|}
\hline
& \multicolumn{3}{|c|}{MAE (Avg. Over Components)} \\
\textit{Pretraining Task:} & SEGNN & SEConv & SETransformer\\
\hline
Formation Energy & 4.67 \textit{(4.82)} & 4.69 \textit{(4.71)} & 4.45 \textit{(4.72)} \\
\hline
Band Gap & 4.66 \textit{(4.82)} & 4.64 \textit{(4.71)} & 4.74 \textit{(4.72)} \\
\hline
\end{tabular}
\caption{Results for predicting the components of the dielectric tensor in experiment 1.}\label{dielectricexp1}
\end{table*}

\subsection{Scalar Elastic Properties}\label{scalar-pred}

The bulk modulus $K$, shear modulus $G$, and Young modulus $E$, are common scalar-valued quantities representing elastic responses in materials.  These values are computed from the elasticity tensor $C$ and compliance tensor $s=C^{-1}$ as the average of the Voigt and Reuss forms as in \cref{scalar-elastic-formulas}. The model used in this work performs worse than MatTen \cite{MatTen} on these scalar tasks, with best results for these scalar predictions including an MAE of $8.41\ GPa$ (compared to $7.37\ GPa$ in \cite{MatTen}) on the $K_H$ test set with the SEConv model, and then $9.04\ GPa$ on the $G_H$ set (compared to $8.38\ GPa$) and $20.06\ GPa$ on $E$ (compared to $20.59\ GPa$) with SEGNN.

Performance in the prediction of these components is presented in \cref{scalar-targets}, where pretraining on scalar target sets is also considered. 
Pretraining on band gap shows no benefit. However, the pretraining task of formation energy did provide slight improvements for the SEConv and SETransformer models. The most notable improvements were observed with the SETransformer model, with test MAE improving from $21.14 \ GPa$ to  $20.35 \ GPa$ on the Young modulus dataset, and an improvement from $9.51\ GPa$ to $8.99 \ GPa$ on the shear modulus dataset.

\begin{table*}[t]\label{scalar-targets}
    \centering
    \begin{tabular}{|l|l|ccc|}
        \hline
        & & \multicolumn{3}{|c|}{MAE} \\
         Model & Pretraining & $K_H$ (GPa) & $G_H$ (GPa) & $E$ (GPa) \\
         \hline
          & None & 8.61 & 9.04 & 20.06 \\
          SEGNN & Band Gap & 9.08 & 9.94 & 22.18 \\
          & Formation Energy & 8.90 & 9.06 & 20.16 \\
         \hline
         & None & 8.41 & 9.37 & 21.61 \\
         SEConv & Band Gap & 8.46 & 9.30 & 20.80\\
         & Formation Energy & 8.42 & 8.81 & 19.56 \\
         \hline
          & None & 9.67 & 9.51 & 21.24 \\
          SETransformer & Band Gap & 10.66 & 10.18 & 23.04 \\
          & Formation Energy & 10.06 & 8.99 & 20.35 \\
         \hline
         MatTen~\cite{MatTen} & None & 7.37 & 8.38 & 20.59 \\
         \hline
    \end{tabular}
    \caption{The MAE of the prediction of the bulk modulus, $K$, shear modulus, $G$, and the young modulus, $E$, derived from the predicted elastic tensor in comparison to MatTen~\cite{MatTen}.}
    \label{scalar_elastic}
\end{table*}

\section{Conclusion}

In a certain sense, the component-wise results and lack of transferability may indicate that the spherical harmonic approach is potentially 'too big' or too expressive to effectively predict crystalline material properties in a way that naturally respects crystalline symmetries. Indeed, $SO(3)$ convolution, as described here, generates a large set of unwanted and consequently discarded combinations of spherical harmonics. 
Future works may incorporate the symmetry of these crystalline systems as in \cite{yan2024equivcrystalprediction}, so that models may naturally predict which tensor components should be zero due to symmetry alone. A simple approach would be to mask results according to crystal system, or build an individual model for each different crystal system with a unique output structure. A more sophisticated approach might be to restrict convolution specifically to the crystal group elements, as opposed to the larger $SO(3)$ group.

The code used in this work can be found in the following GitHub repository: 
\hyperlink{https://github.com/qmatyanlab/Transfer-Learning-For-Complex-Crystal-Properties}{https://github.com/qmatyanlab/Transfer-Learning-For-Complex-Crystal-Properties}.
           
\printbibliography

\appendix
\onecolumn

\section{Constructing $SO(3)$-Invariant Subspaces of Tensors}\label{youngoverview}

An invariant subspace of a tensor is a subspace which is closed under the action of some group of transformations. That is, invariant and disjoint subspaces of a tensor shouldn't 'mix' at all under the corresponding group transformations. 

Invariants under $SO(3)$ of an arbitrary tensor $T$ may be constructed by way of Young symmetrizers, and subsequent contractions with the metric tensor $g_{ij}$ or the fully antisymmetric tensor $\epsilon_{ijk}$. Note that these three methods correspond to the decomposition of a tensor with respect to the general linear group (by way of Young symmetrizers), the orthogonal group (contractions with $g_{ij}$), and the special linear group (contractions with $\epsilon_{ijk}$), respectively. 

Since $SO(3) =   SL(3)\bigcap O(3) \subset GL(3) $, to construct invariants of $SO$, we generally form invariant subspaces under $GL$ by way of Young symmetrizers first, and then construct invariants under $O$ and/or $SL$ of these $GL$ invariant subspaces. This often results in several different harmonic subspaces of the same order $\ell$  in the harmonic decomposition of a tensor.
Also, note that the $GL$ decompositions for tensors of rank greater than two are not, in general, unique.

Below, we introduce the tools used in the construction of such invariant subspaces and demonstrate their invariance. These tools are then applied in three cases of tensor spaces with importance in the field of materials science: corresponding to the space of dielectric tensors, piezoelectric response tensors, and elasticity tensors.

\subsection{$GL$ Decompositions}
By way of the Schur-Weyl duality, the invariant subspaces of a tensor under the general linear group $GL$ determine the invariant subspaces of the tensor under the symmetric group $S_n$, where symmetric group elements act as permutations on tensor indices.

Under the symmetric group, the invariant subspaces of a tensor are described by way of standard Young tableaux, and then constructed from an arbitrary tensor by means of the corresponding Young symmetrizers. Below, these concepts are briefly introduced for the purpose of $GL$ decompositions of tensors over $\mathbb{C}^n$. Note that these decompositions are in general not unique for tensors of rank $>2$, so some conventions must be adopted. In the present work we adopt the convention of antisymmetrization before symmetrization in the construction of Young symmetrizers.

\subsubsection{Young Diagrams, Tableaux and Symmetrizers}

Young diagrams of order $n$ are left-justified arrangements of boxes into $k$ rows stacked vertically in non-increasing order. 
A Young diagram is said to be of some shape $\lambda:(\lambda_1,\lambda_2,...,\lambda_k)$, where $\lambda_i$ refers to the depth of row $i$ and $\lambda_{i+1}\leq\lambda_i\leq\lambda_{i-1}$. 

\begin{figure}[!h]
\centering
\includegraphics[scale=0.6]{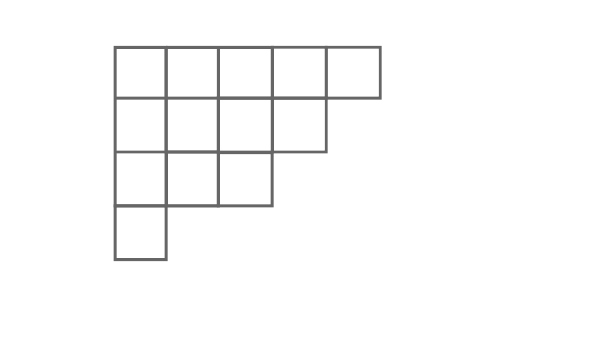}
\caption{Example Young diagram of shape $(5,4,3,1)$.}
\end{figure}

We can form a set of Young tableaux from diagrams by filling in the boxes with indices from an ordered set $\lbrace x_1,x_2,...,x_k\rbrace$ corresponding to tensor components as $T^{x_1x_2...x_k}$.  A standard tableu is one filled with indices $x_i$ (without repeats) with entries increasing in index $i$ down each column and across (to the right) rows.
\begin{figure}\centering
\includegraphics[scale=0.5]{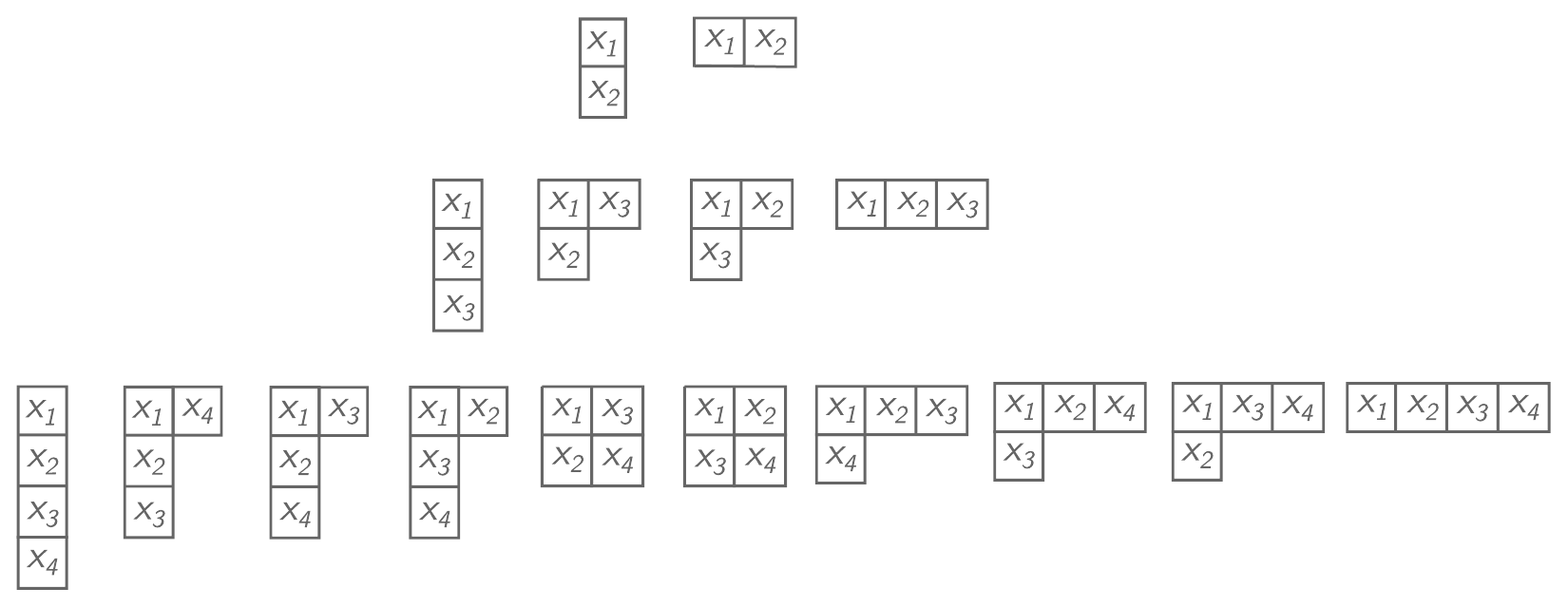}
\caption{Standard Young tableaux for diagrams of at most 4 boxes.}
\end{figure}

Each of these standard tableaux correspond to an invariant subspace under $S_k$, and further $GL_n$, by way of the Schur-Weyl Duality. From these tableau, we may construct so-called \textit{Young Symmetrizers}, which project tensors onto their corresponding $S_k$-invariant subspace. 

A Young symmetrizer $P_{\lambda}$ corresponding to a tableau $\lambda$ is composed of a compound set of symmetrizing operations $s_{\lambda}$ and antisymmetrizing operations $a_{\lambda}$, scaled by an overall normalization constant $C_{\lambda}$:
$$
P_{\lambda} = \mathcal{C}_{\lambda}s_{\lambda}a_{\lambda}
$$
Where here, we adopt the convention of anti-symmetrization before symmetrization, following that in Itin \cite{itin-rank3}.

The symmetrizing operator $s_{\lambda}$ for a diagram $\lambda$ is composed of a product of symmetrizers $\mathcal{S}(\mathcal{I})$, where $\mathcal{I}$ ranges over all subsets of indices corresponding to some vertically stacked set of indices in tableau $\lambda$, i.e.:
$$
s_{\lambda} = \prod_{\mathcal{I}\in \text{Cols}(\lambda)}\mathcal{S}(\mathcal{I})
$$
where $\text{Cols}(\lambda)$ represents the set of disjoint subsets of indices down each column, and symmetrizers $\mathcal{S}$ are defined to act on tensors $T$ component-wise as:
$$
\big[\mathcal{S}(\mathcal{I})T\big]_{ijk...}= \sum_{\sigma_{\mathcal{I}}}T_{\sigma_{\mathcal{I}}(ijk...)}
$$
where $\sigma_{\mathcal{I}}$ are permutations of index subset $\mathcal{I}$.

Similarly, the antisymmetrizing operator $a_{\lambda}$ can be constructed as a product of antisymmetrizers:
$$
a_{\lambda} = \prod_{\mathcal{I}\in \text{Rows}(\lambda)}\mathcal{A}(\mathcal{I})
$$
where $\text{Rows}(\lambda)$ represents the set of  disjoint subsets of indices across each entire row, and antisymmetrizers $\mathcal{A}$ are defined as:
$$
\big[\mathcal{A}(\mathcal{I})T\big]_{ijk...}= \sum_{\sigma_{\mathcal{I}}}\text{sgn}(\sigma_{\mathcal{I}})T_{\sigma_{\mathcal{I}}(ijk...)}
$$
where, again, $\sigma_{\mathcal{I}}$ range over all permutations of index subset $\mathcal{I}$.

The normalization constant $C_{\lambda}$ may be derived from the shape of the underlying Young diagram according to the \textit{hook-length formula}, given below, where $\text{hook}(\alpha,\beta)$ returns the number of boxes crossed by a hook coming up (from below) column $\beta$ and out of the diagram to the right in row $\alpha$.
$$
C_{\lambda}= \prod_{(\alpha,\beta)\in \lambda}\frac{1}{\text{hook}(\alpha,\beta)}
$$
Furthermore, the number of independent components $N_{\lambda}$ of a tensor subspace corresponding to some Young tableau with shape $\lambda$ may also be derived from the diagram by a related hook-length formula:
$$
N_{\lambda}= \prod_{(\alpha,\beta)\in \lambda}\frac{n-\alpha+\beta}{\text{hook}(\alpha,\beta)}
$$
where $n$ is the dimension of the vector space forming the tensor space. 

Thus, the standard Young tableaux of $k$ boxes can be used to decompose an arbitrary tensor space into a set of $GL$ invariant subspaces with known symmetries (under permutation of indices) by way of corresponding Young symmetrizers. These known symmetries will be relevant in the further decomposition of these subspaces by way of contractions with the metric tensor, and the fully antisymmetric tensor, discussed below.
\begin{figure}\centering
\includegraphics[scale=0.6]{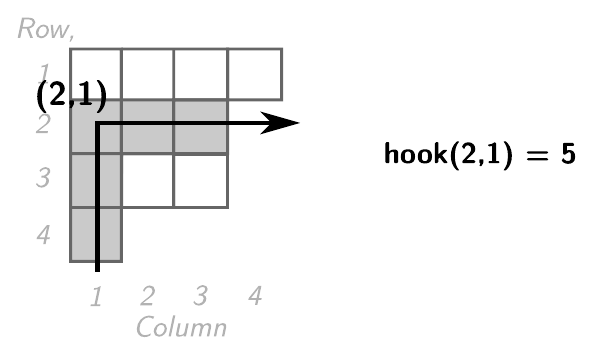}
\caption{Example value of $\text{hook}(2,1)$ for a diagram of shape $\lambda:(4,3,3,1)$. Note that it's corresponding normalization constant is $C_{\lambda}=1/33600$.}
\end{figure}
\subsection{$SL$ Decompositions}
The special linear group $SL$ of transformations is defined as the subset of invertible linear transformations with determinant equal to positive one. Under $SL$, orientation and volume are preserved, where volume is defined as the contraction of a tensor with the fully antisymmetric tensor $\epsilon_{ijk}$, which transforms under $R\in GL(3)$ as:
$$
\epsilon_{x_1x_2x_3} \rightarrow\epsilon_{x_1'x_2'x_3'} =\text{det}(R) R_{x_1'}^{x_1} R_{x_2'}^{x_2}R_{x_3'}^{x_3}\epsilon_{x_1x_2x_3} 
$$
For $R\in SL$ then, we have $\text{det}(R)=1$, allowing us to form various invariants of tensors by way of contraction (or partial contraction) with $\epsilon$. 

For example, for a rank-three tensor $T$, the volume may be defined component-wise as:
$$
V = \epsilon_{ijk}T^{ijk}
$$
This $V$ is then clearly invariant under $SO$ since for any transformation $R\in SO(3)$, we have:
$$
V\rightarrow\epsilon_{i'j'k'}T^{i'j'k'} =  R_{i'}^{i} R_{j'}^{j}R_{k'}^{k}\epsilon_{ijk}R_{i}^{i'} R_{j}^{j'}R_{k}^{k'}T^{ijk} =\epsilon_{ijk}T^{ijk}
$$

Note that contractions with $\epsilon$ along fully symmetric sets of indices will always vanish; a fact useful when considering particular $GL$ invariant subspaces. Explicitly,
$$
\epsilon_{ijk}T_{..(ijk)..} = 0
$$

\subsection{$O$ Decompositions}
The orthogonal group $O$ of transformations is defined as the subset of invertible linear transformations satisfying $R^TR=1$. Under $O$, we may define an inner product between vectors $<\cdot , \cdot >$ by means of a metric tensor $g_{ij}$ ($\delta_{ij}$ in Euclidean space) , which transforms as a second rank tensor under some transformation $R$ as:
$$
g_{x_1x_2}\rightarrow g_{x'_1x_2'} = R_{x_1'}^{x_1} R_{x_2'}^{x_2}g_{x_1x_2}
$$
The inner product of vectors $u,v$ is defined component-wise as:
$$
<u,v> = u^i g_{ij}v^j
$$
which is then clearly invariant under a transformation $R\in O(3)$ as:
$$
<u',v'> = R_{i}^{i'}u^i R_{j'}^{j} R_{j'}^{j} g_{ij}R_{j}^{j'}v^j =u^i g_{ij}v^j =<u,v> 
$$
 
Furthermore, we may construct different invariants for different rank tensor spaces. For example, we may define $O$ invariant scalar-valued traces  $Tr(\cdot)$ of second rank tensors $M$ by means of the metric tensor, defined component wise as
$$
Tr(M) = g_{ij}M^{ij}
$$
and we may form three $O$ invariant trace vectors for third rank tensors $T$, defined component wise as:
$$
\nu^{k} = g_{ij}T^{ijk},\quad \mu^{j} = g_{ki}T^{ijk},\quad u^{i} = g_{jk}T^{ijk}
$$
That is, with the metric tensor $g_{ij}$ available, as is the case in $SO(3)\subset O(3)$, we may decompose a rank-$n$ tensor into a sets of rank-$(n-2),(n-4),...,(0$ or $1)$ tensors by taking sequential traces.


The trace-part of a tensor may then be reconstructed by taking tensor products of metric tensors and trace vectors.

A point which is relevant to $GL$ invariant subspaces is that contractions with respect to $g$ under fully antisymmetric pairs of indices will always vanish. That is,
$$
g_{ij}T_{..[ij]..} = 0
$$

\subsection*{SO(3) Decompositions}

In $SO(3)$, we have all of the above tools (Young diagrams, $g_{ij}$, and $\epsilon_{ijk}$) available. As such, we may decompose a tensor into a set of $SO(3)$ invariant subspaces by their use, and all will be valid. Unfortunately, for tensors of rank$>2$, there is not, in general, a unique and irreducible set of invariant $SO(3)$ spaces. However, we may always decompose an arbitrary tensor into a set of (not necessarily unique) irreducible $SO(3)$-invariant symmetric tensor subspaces using the tools given above.

\section{Full Decompositions of Tensors}\label{decomps}
Below, we give the $SO(3)$ decomposition of the three tensors of interest here. Namely, the rank-two dielectric tensor $\epsilon$, the rank-three piezoelecrtic strain tensor $d$, and the rank-four elastic tensor $C$. Each begins with the Young diagrams for the total rank tensor space and then uses the symmetries of the respective tensor to restrict down to the relevant Young tableaux. These are used to form symmetrizers and then further decomposed with contractions with the metric tensor $g$ and the levi-civita tensor $\epsilon$ until all that remains are a set of fully symmetric sub-tensors.
\subsection{Dielectric Tensor Decomposition}
The decomposition of the dielectric tensor is essentially trivial since it is already symmetric. However, we give a detailed overview here for instructive purposes.

The GL decomposition of a rank-two tensor space is the usual symmetric-antisymmetric decomposition of a matrix $M$.
$$
M_{ij} = S_{ij} + A_{ij}\quad
$$
\begin{center}
with:
\begin{align*}
S_{ij}&=\frac{1}{2}(M_{ij}+M_{ji})\quad \Rightarrow\quad S_{ij}=S_{ji}\\
A_{ij}&=\frac{1}{2}(M_{ij}-M_{ji})\quad \Rightarrow\quad A_{ij}=-A_{ji}\\
\end{align*}
which correspond to the Young diagrams below:

\includegraphics[scale=0.7]{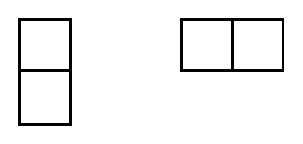}.
\end{center}
Since the dielectric tensor $\epsilon$ is symmetric, we are left then only with the unique tableaux below, corresponding to the fully symmetric part $S$.
\begin{center}
\includegraphics[scale=0.7]{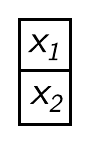}
\end{center}
We can then further decompose this rank-two symmetric into a rank-zero space and a rank-two space by taking the rank-0 trace $t = g_{ij}\epsilon_{ij}$ and then forming the rank-2 traceless residue $R_{ij}=\epsilon_{ij}-t\delta_{ij}$.
\subsection{Piezoelectric Tensor Decomposition}
The Young diagrams for a rank-three tensor are as follows:
\begin{center}
\includegraphics[scale=0.7]{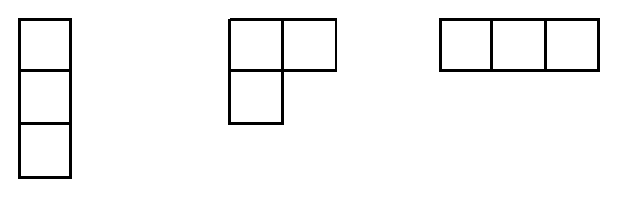}
\end{center}
according to the symmetry $d_{ijk}=d_{ikj}$, we can again see that all Young tableaux but the following will disappear, since the rest are asymmetric in the last two indices.
\begin{center}
\includegraphics[scale=0.7]{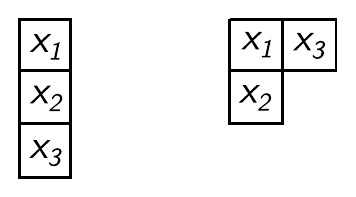}
\end{center}
where the totally symmetric component tensor here we define as $S$ and the mixed symmetry component tensor we define $A$, so that we have the Young decomposition:
$$
d_{ijk} = S_{ijk}+A_{ijk}
$$
defined component-wise in terms of strain components:
\begin{align*}
S_{ijk}&=\frac{1}{3}\big(d_{ijk}+d_{ikj}+d_{kji}\big)\\
A_{ijk}&=\frac{1}{3}\big(2d_{ijk}-d_{ikj}-d_{kji}\big)\\
\end{align*}

The fully symmetric part $S$ is of an adequate form for harmonic decomposition, with those relations given in the next section.

The mixed symmetry part $A$ however requires further decomposition with respect to $SO(3)$, so that we have a set of symmetric tensors describing it. It's 8 independent components can be described by a $5\oplus 3$ dimensional space consisting of a symmetric rank-2 tensor and a trace vector.

The trace vector $v^i$, which corresponds to $A$'s 3-dimensional $SO(3)$ invariant subspace, can be formed from the contraction of the metric tensor $g$ along $A$'s first and second indices. That is, we define:
$$
v^i =g_{jk}A^{ijk}
$$
Note that this choice is somewhat arbitrary, since we could define the trace part to correspond to the contraction along the first and third indices, or the second and third. However, it can be shown that for the mixed symmetry of $A$, these two potential trace vectors are linearly dependent (related by an overall factor of $1$ and $-2$, respectively).

Rather simply then, we can reconstruct a third-rank tensor $V$, corresponding to this rank-one invariant subspace, by defining $V$ component-wise as:
$$
V_{ijk} =\frac{1}{4}\big[v_i g_{jk}+ v_j g_{ik} - 2 v_j g_{ik}\big]
$$
The rank-2 invariant subspace of $A$ then may be constructed by symmetrizing the partial contraction with $\epsilon$ along the first and second indices (the anti-symmetric pair). Note that the antisymmetric part of this partial contraction corresponds to the trace vector space accounted for here by $u$. Explicitly, we define:
$$
b_{ij}=\frac{1}{2}\big(\epsilon_{i}^{mk}A_{mkj}+\epsilon_{j}^{mk}A_{mki}\big)
$$
which is a traceless symmetric rank-2 tensor. And from which we may reconstruct the corresponding rank-3 tensor $B$, defined by:
$$
B_{ijk}=\frac{1}{3}\big[ \epsilon_{ik}^p b_{pj}+\epsilon_{ij}^p b_{pk}\big]
$$
Thus, we may provide a total harmonic decomposition for $A$ in terms of the invariant subspaces of the rank-1 $u$ and the rank-2 $b$, such that:
$$
A_{ijk} = V_{ijk} + B_{ijk}
$$
And so, as mentioned above, the symmetric part $S$ is readily decomposed in the spherical bases into $\mathcal{H}^{(1)}\oplus \mathcal{H}^{(3)}$. And then the mixed-symmetry part inhabits the space $\mathcal{H}^{(1)}\oplus \mathcal{H}^{(2)}$.

\subsection{Elastic Tensor Decomposition}\label{elastic_decomp_app}

If we examine the Young diagrams of a fourth rank tensor, as below: 

\begin{center}
\includegraphics[scale=0.7]{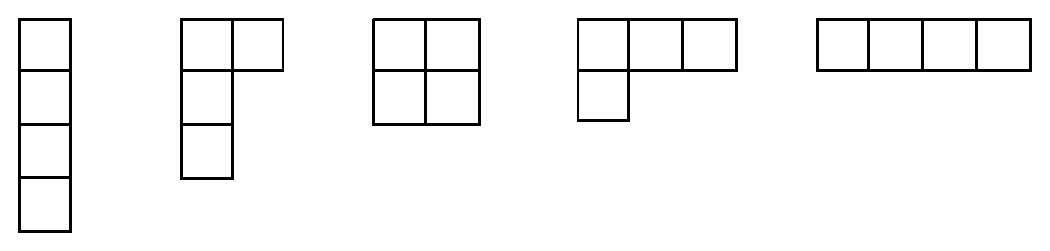}
\end{center}
We can immediately notice that the symmetries of $C$ require that all but the totally symmetric tensor $S$ and one mixed symmetry tensor $A$ must vanish. $S$ has 15 independent components, and $A$ has 6. Furthermore, $A$ has corresponding symmetrizer $\mathcal{C}S(x_1x_2)S(x_3x_4)A(x_1x_3)A(x_2x_4)$. Thus, $A$ is exactly the tensor symmetric under permutation of $i,j$ and $k,l$ but antisymmetric under exchanges $i,k$ and $j,l$.
\begin{center}
\includegraphics[scale=0.7]{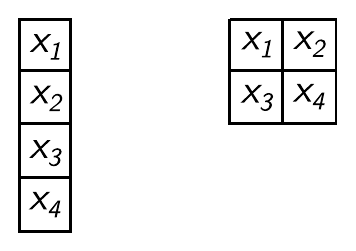}
\end{center}
Note that this mixed-symmetry subspace $A$ corresponds to Backus' \cite{backus1970geometrical} asymmetric tensor $A$. With $S$ and $A$ being defined component-wise as:
$$
S_{ijkl}=\frac{1}{3}\big( C_{ijkl} + C_{ikjl} + C_{klij} \big)
$$
$$
A_{ijkl} = \frac{1}{3}\big( 2C_{ijkl} -C_{ikjl} -C_{klij}  \big)
$$
The fully symmetric part can of course be converted to spherical harmonic components via a Clebsch-Gordon expansion. The mixed symmetry $A$, however requires further decomposition. 

All six components can be described by the symmetric (but not traceless) tensor $t$, defined as the double partial contraction of $A$ with the totally antisymmetric tensor $\epsilon$ as:
$$
t_{ij} = \epsilon_{i}^{mk}\epsilon_{j}^{nl}A_{mnkl}
$$
\begin{center}
which can be reconstructed as the subtensor $N$ as:
\end{center}
$$
N_{ijkl}= \frac{1}{2}\big(\epsilon^{}\epsilon^{} - \epsilon^{} \epsilon^{} \big)t_{mn}
$$
This tensor $t$ then has a harmonic decomposition according to the rank-two Clebsch-Gordon transformation between the $J_z$ basis and the harmonic basis $y_{\ell}^m$.

\section{Real Spherical Harmonics}\label{realsph}
The (complex) spherical harmonics $Y_{\ell}^m$ can be transformed into a set of real spherical harmonics $Y_{\ell m}$ according the following relations:
$$
Y_{\ell m }=
\begin{cases}
\frac{i}{\sqrt{2}}\big(Y_{\ell}^{-|m|} - (-1)^m Y_{\ell}^{|m|} \big):\ \ m<0\\
\quad\quad\quad\quad Y_{\ell}^0:\quad\quad\quad\quad\quad\ \ \  \ m=0\\
\frac{1}{\sqrt{2}}\big(Y_{\ell}^{-|m|} + (-1)^m Y_{\ell}^{|m|} \big):\ \ m>0\\
\end{cases}
$$
(of course, this choice is not entirely unique, with the choice made here assuming a Condon-Shortley phase included in the definition of $Y_{\ell}^m$).

\section{Derivation of Scalar Elastic Properties from Elastic Tensor}\label{scalar-elastic-formulas}
\begin{equation*}
    K_V = [(C_{11}+C_{22}+C_{33}) + 2(C_{12}+C_{23}+C_{31})]/9
\end{equation*}
\begin{equation*}
    K_R = 1/[(s_{11}+s_{22}+s_{33}) + 2(s_{12}+s_{23}+s_{31})]
\end{equation*}
\begin{equation*}
    K_H = (K_V+K_R)/2
\end{equation*}
\begin{equation*}    
    G_V = [(C_{11}+C_{22}+C_{33}) - (C_{12}+C_{23}+C_{31}) + 3(C_{44} + C_{55} + C_{66})]/15
\end{equation*}
\begin{equation*}
    G_R = 15/[4(s_{11}+s_{22}+s_{33}) - 4(s_{12}+s_{23}+s_{31}) + 3(s_{44} + s_{55} + s_{66})]
\end{equation*}
\begin{equation*}
    G_H = (G_V+G_R)/2
\end{equation*}
\begin{equation*}
    E = 9K_HG_H/(3K_H+G_H)
\end{equation*}

\newpage

\begin{table*}[!t]
\section{Table of Coefficients for $J_z\rightarrow y_{\ell}^m$ Transformation}\label{cgcoef}
\begin{tabular}{c|ccccccc}
$\mathbf{Rank\ 2}$ \\
\hline
& $a_{00}$ & $a_{+-}$ & $a_{0\pm}$ & $a_{\pm\pm}$ \\
$y_0^0$  & $1$ & $-2$ \\
$y_2^0$  & $1$ & $1$ \\
$y_2^{\pm 1}$  &  & & $\sqrt{3}$ \\
$y_2^{\pm 2}$  &  & & & $\sqrt{\frac{3}{2}}$
\\
\\
\\
$\mathbf{Rank\ 3}$\\
\hline
 & $a_{000}$ & $a_{0+-}$ & $a_{00\pm}$ & $a_{+-\pm}$  & $a_{0\pm\pm}$ & $a_{\pm\pm\pm}$ \\
$y_1^0$  & $3$ & $-6$ \\
$y_3^0$  & $\frac{15}{7}$ & $-\frac{5}{7}$ \\
$y_1^{\pm 1}$  &  & & $\sqrt{\frac{3}{2}}$ & $-2\sqrt{\frac{3}{2}}$\\
$y_3^{\pm 1}$  &  & & $2\sqrt{\frac{3}{2}}$ & $\sqrt{\frac{3}{2}}$\\
$y_3^{\pm 2}$  &  & & & & $\sqrt{\frac{15}{2}}$\\
$y_3^{\pm 3}$  &  & & & & & $\sqrt{\frac{5}{2}}$\\
\\
\\
\\
\textbf{Elastic}\\
(Rank 4) \\
 \hline 
$(m=0)$& $a_{0000}$ & $a_{00+-}$ & $a_{0+0-}$ & $a_{+-+-}$  & $a_{++--}$ \\
$y_0^{0}$  & $1$ & $2$ &$-6$ & 1 & 3 \\
$y_2^0$  & 1 & $2$ & $-3$ & 1 & $-3$\\
$y_4^0$ & 1 & $2$ & $4$ & 1 & $\frac{1}{2}$\\
 \\
$(m=1,2)$& $a_{000\pm}$ & $a_{+- 0 \pm}$  & $a_{\mp 0 \pm\pm}$ & $a_{00\pm\pm}$ & $a_{+- \pm \pm}$  & $a_{\mp \pm \mp\pm}$ \\
$y_2^{\pm 1}$  & $\sqrt{3}$ & $\sqrt{3}$ &  $-3\sqrt{3}$\\
$y_4^{\pm 1}$  & $2\sqrt{\frac{5}{2}}$ & $2\sqrt{\frac{5}{2}}$ & $\sqrt{\frac{5}{2}}$  \\
$y_2^{\pm 2}$  & & & & $-2\sqrt{\frac{3}{2}}$ & $-2\sqrt{\frac{3}{2}}$ & $3\sqrt{\frac{3}{2}}$  \\
$y_4^{\pm 2}$ & & & & $\sqrt{\frac{5}{2}}$ & $\sqrt{\frac{5}{2}}$ & $2\sqrt{\frac{5}{2}}$  \\
\\
$(m=3,4)$& $a_{\pm\pm\pm 0}$ & $a_{\pm\pm\pm\pm}$  \\
$y_4^{\pm 3}$  & $\sqrt{\frac{35}{2}}$ &  \\
$y_4^{\pm 4}$  &  & $\frac{1}{2}\sqrt{\frac{35}{2}}$\\ 
\end{tabular}
\caption{Coefficients for transformation between spherical basis components $a_{\alpha\beta\gamma\delta}$ and harmonic components $y^m_l$ for a rank-2,3,4 symmetric tensor. Note that the rank-4 is only relevant for symmetric components of elastic tensors and correspond to Mochizuki's transformation \cite{mochizuki1988spherical} between the $J_z$ basis and their symmetric tensor $s$.}
\end{table*}

\newpage
\section{Distribution of Data}\label{data-dist}

There were 3292 piezoelectric tensors, 7273 dielectric tensors, and 10286 elasticity tensors in the datasets used. This is after pruning all those available on Materials Project \cite{matproj} to those with spherical harmonic components below some high-end cutoff, here 5000.

\begin{sidewaystable}


\large

\begin{tabular}{|c c c c c c|}
 \hline
   \multicolumn{6}{|c|}{Elasticity Components (Mean $\pm$ STD in $GPa$)} \\
 \hline     177.79 $\pm$ 134.19 &  72.27 $\pm$ 54.25   &  69.36 $\pm$ 53.86   &  0.02 $\pm$ 3.89   &  0.05 $\pm$ 4.12   &  0.02 $\pm$ 1.83  \\
  72.27 $\pm$ 54.25   &  178.16 $\pm$ 134.71 &  69.32 $\pm$ 53.86   &  -0.01 $\pm$ 4.18  &  0.08 $\pm$ 2.89   &  0.02 $\pm$ 2.02  \\
  69.36 $\pm$ 53.86   &  69.32 $\pm$ 53.86   &  175.33 $\pm$ 142.22 &  -0.02 $\pm$ 1.74  &  0.06 $\pm$ 3.91   &  -0.0 $\pm$ 1.06  \\
  0.02 $\pm$ 3.89     &  -0.01 $\pm$ 4.18    &  -0.02 $\pm$ 1.74    &  53.37 $\pm$ 47.03 &  0.0 $\pm$ 0.68    &  0.08 $\pm$ 2.88  \\
  0.05 $\pm$ 4.12     &  0.08 $\pm$ 2.89     &  0.06 $\pm$ 3.91     &  0.0 $\pm$ 0.68    &  53.86 $\pm$ 47.23 &  0.04 $\pm$ 3.82  \\
  0.02 $\pm$ 1.83     &  0.02 $\pm$ 2.02     &  -0.0 $\pm$ 1.06     &  0.08 $\pm$ 2.88   &  0.04 $\pm$ 3.82   &  55.74 $\pm$ 47.23\\
 \hline
\end{tabular}

\medskip

\begin{tabular}{|c c c c c c|}
 \hline
   \multicolumn{6}{|c|}{Piezoelectric Components (Mean $\pm$ STD in $C/m^2$)} \\
 \hline 
  0.007 $\pm$ 1.382    &  -0.0149 $\pm$ 0.6281 &  -0.0207 $\pm$ 0.5903 &  0.0124 $\pm$ 0.4783 &  0.0186 $\pm$ 2.5957  &  -0.0163 $\pm$ 0.7636\\
  -0.0084 $\pm$ 1.0081 &  0.0384 $\pm$ 1.0524  &  0.0088 $\pm$ 0.6786  &  0.0515 $\pm$ 2.2186 &  0.0078 $\pm$ 0.408   &  0.0048 $\pm$ 0.4732 \\
  -0.0099 $\pm$ 1.2187 &  -0.0162 $\pm$ 1.0639 &  -0.003 $\pm$ 1.0599  &  0.0002 $\pm$ 0.3991 &  -0.0084 $\pm$ 0.4473 &  0.0072 $\pm$ 0.7069  \\
 \hline
\end{tabular}

\medskip

\begin{tabular}{|c c c|}
 \hline
   \multicolumn{3}{|c|}{Dielectric Components (Mean $\pm$ STD)} \\
 \hline 
   21.89 $\pm$ 72.55 &  0.02 $\pm$ 4.46   &  0.38 $\pm$ 18.24 \\
  0.02 $\pm$ 4.46   &  21.62 $\pm$ 71.08 &  -0.01 $\pm$ 4.04 \\
  0.38 $\pm$ 18.24  &  -0.01 $\pm$ 4.04  &  21.57 $\pm$ 85.67 \\
 \hline
\end{tabular}

\caption{Component-wise mean and standard deviation for tensorial datasets used in this paper.}\label{datavariance}
\end{sidewaystable}

\newpage
\section{Elastic Tensor Component-wise MAE}

\subsection{Elastic Tensor Prediction from Band Gap}

\begin{figure}[h]
    \centering
    \begin{subfigure}{0.45\textwidth}
        \includegraphics[width=\textwidth]{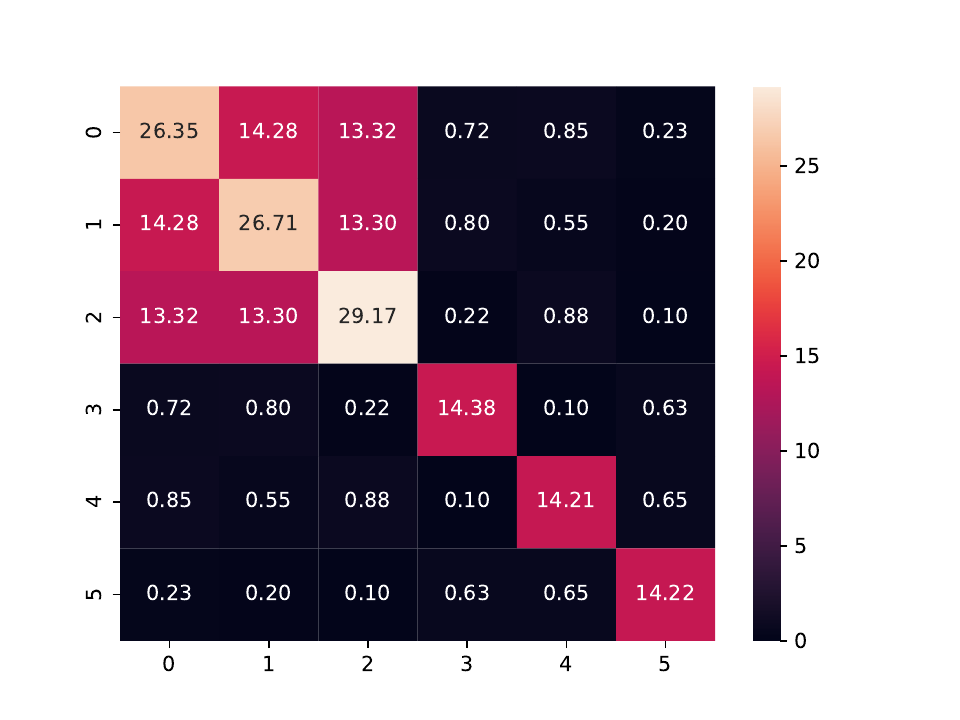}
        \caption{SEConv Component-wise Performance}
    \end{subfigure}
    \begin{subfigure}{0.45\textwidth}
        \includegraphics[width=\textwidth]{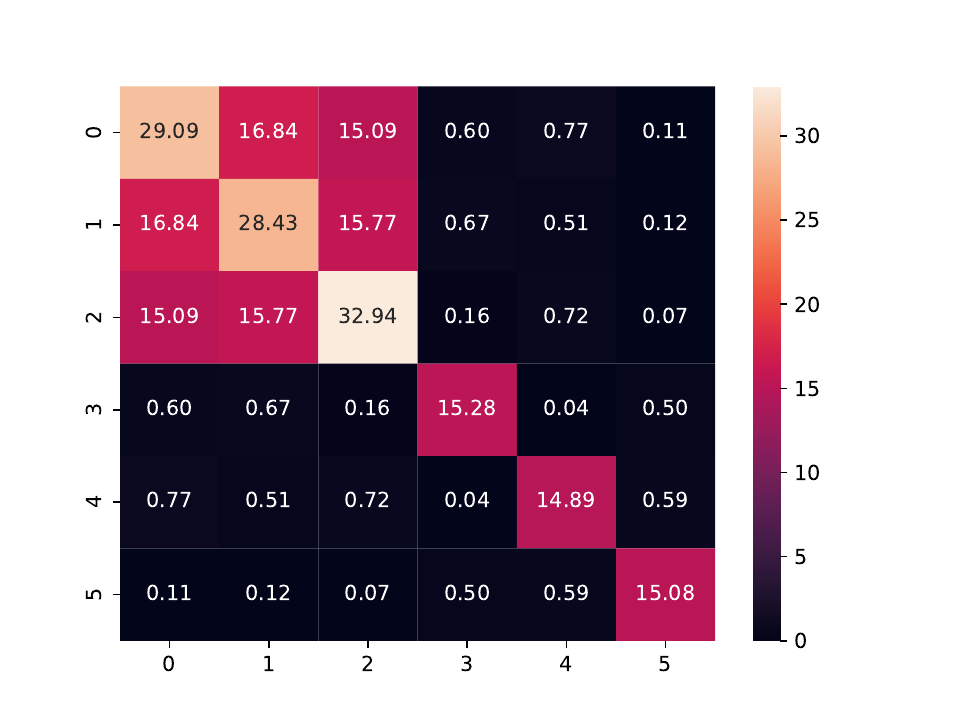}
        \caption{SETransformer Component-wise Performance}
    \end{subfigure}
    \begin{subfigure}{0.45\textwidth}
        \includegraphics[width=\textwidth]{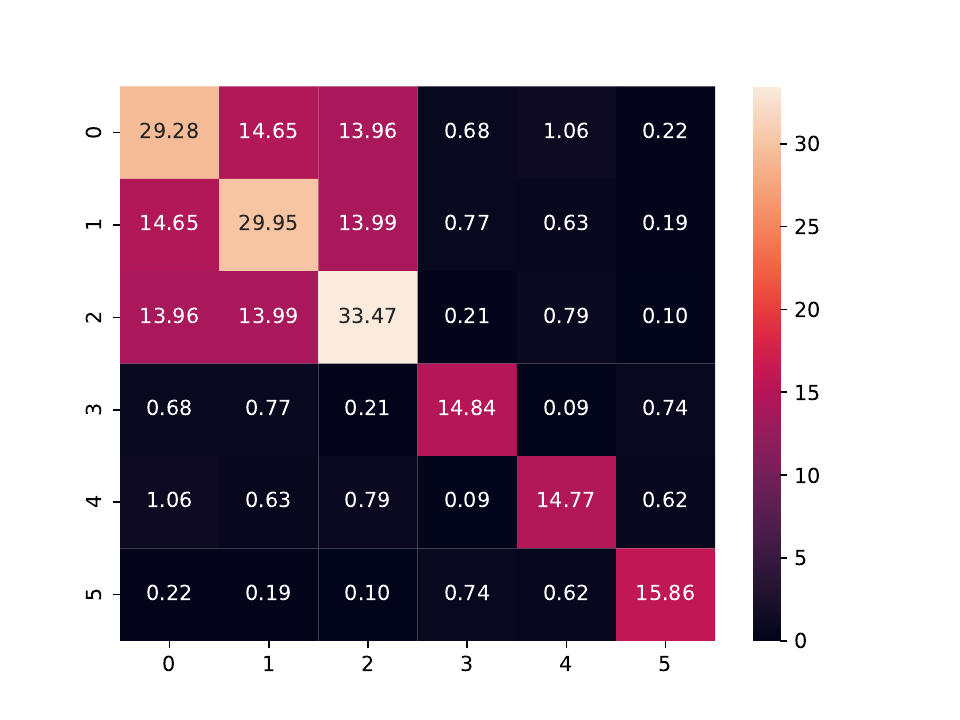}
        \caption{SEGNN Component-wise Performance}
    \end{subfigure}
    \label{BG-ET}
\end{figure}

\newpage
\subsection{Elastic Tensor Prediction from Formation Energy}

\begin{figure}[h]
    \centering
    \begin{subfigure}{0.45\textwidth}
        \includegraphics[width=\textwidth]{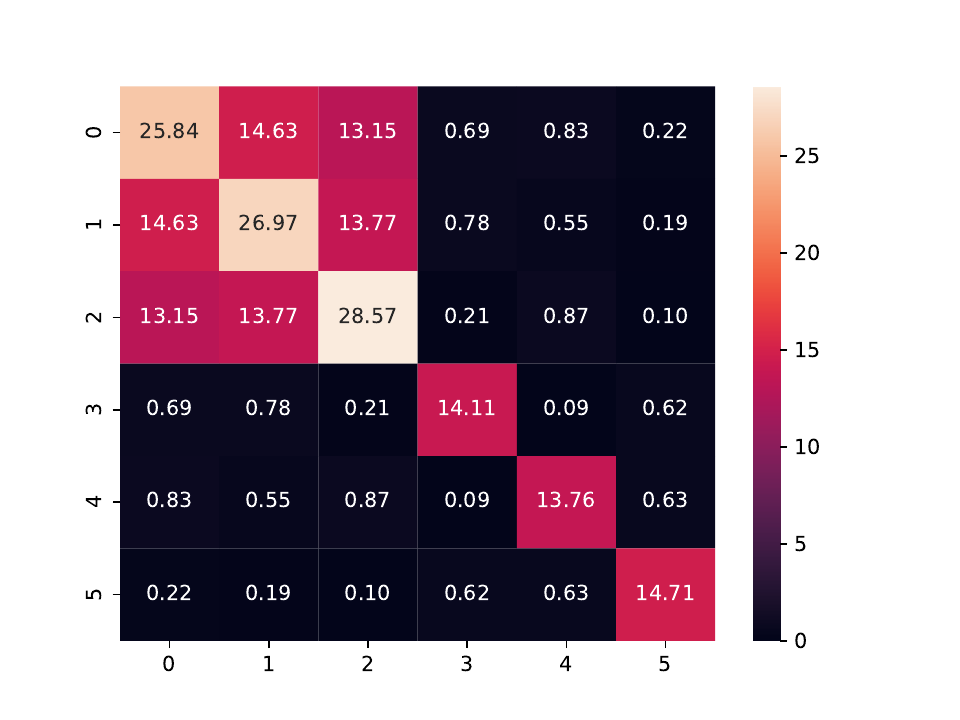}
        \caption{SEConv Component-wise Performance}
    \end{subfigure}
    \begin{subfigure}{0.45\textwidth}
        \includegraphics[width=\textwidth]{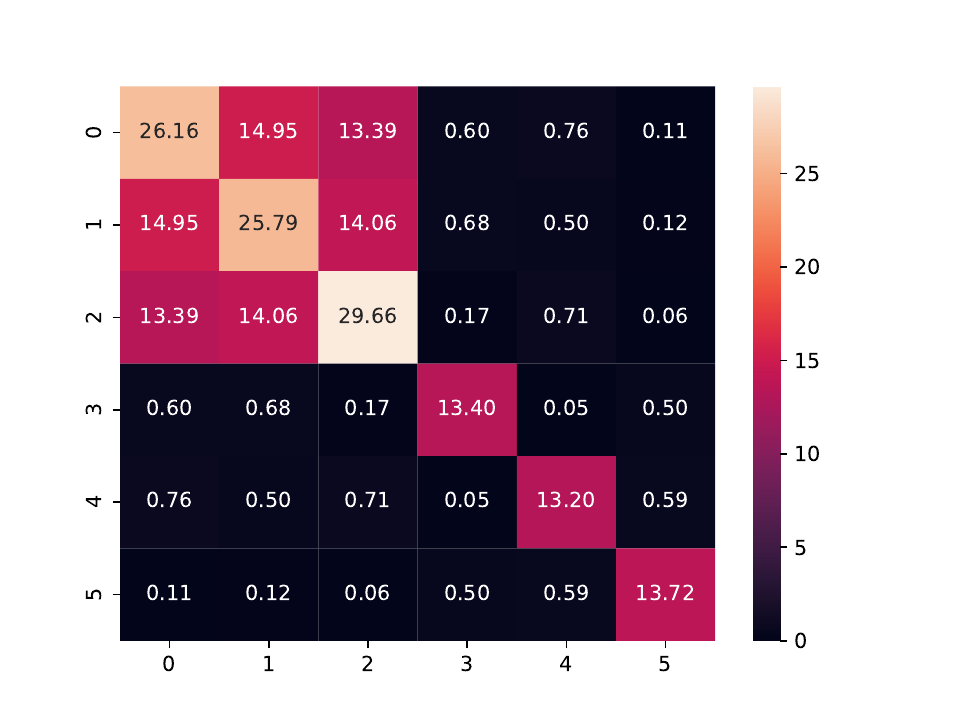}
        \caption{SETransformer Component-wise Performance}
    \end{subfigure}
    \begin{subfigure}{0.45\textwidth}
        \includegraphics[width=\textwidth]{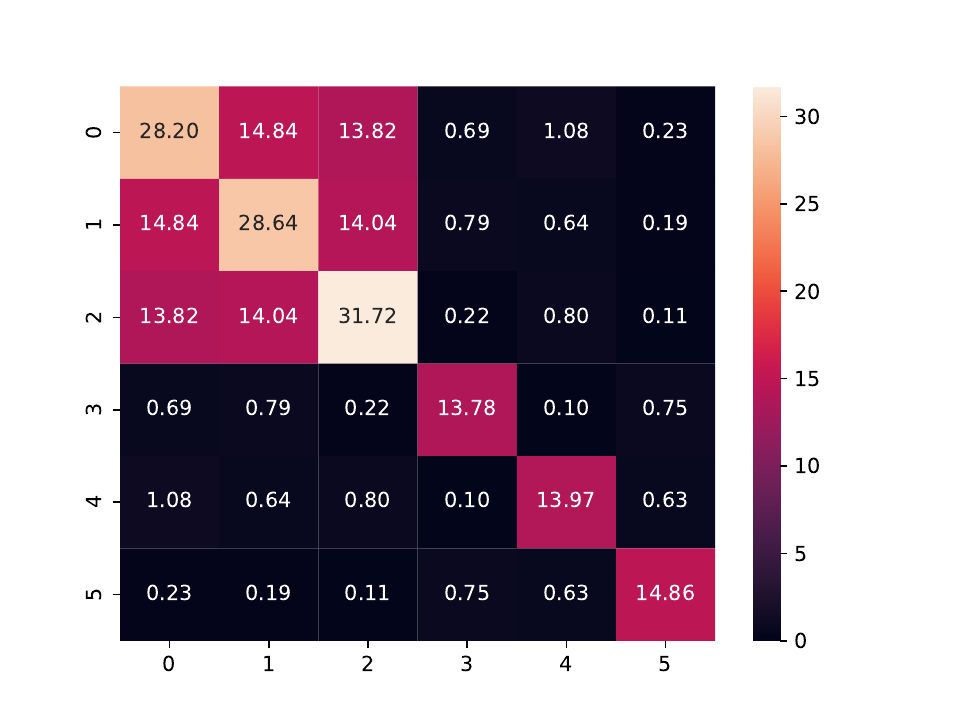}
        \caption{SEGNN Component-wise Performance}
    \end{subfigure}
    \label{FE-ET}
\end{figure}

\newpage
\section{Dielectric Tensor Component-wise MAE}

\subsection{Dielectric Tensor Prediction from Band Gap}

\begin{figure}[h]
    \centering
    \begin{subfigure}{0.45\textwidth}
        \includegraphics[width=\textwidth]{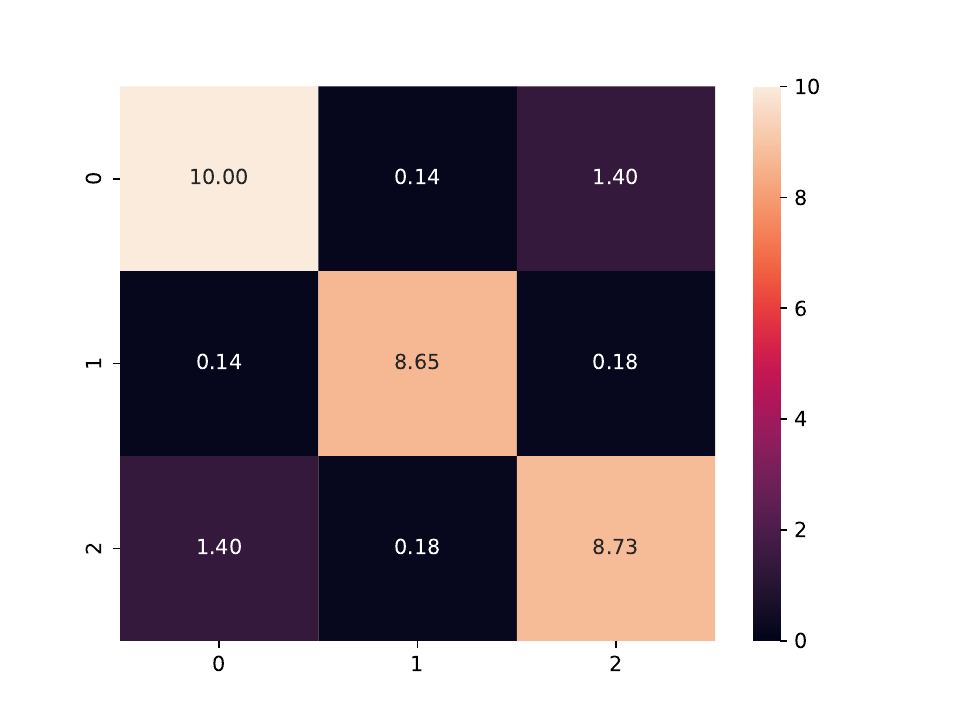}
        \caption{SEConv Component-wise Performance}
    \end{subfigure}
    \begin{subfigure}{0.45\textwidth}
        \includegraphics[width=\textwidth]{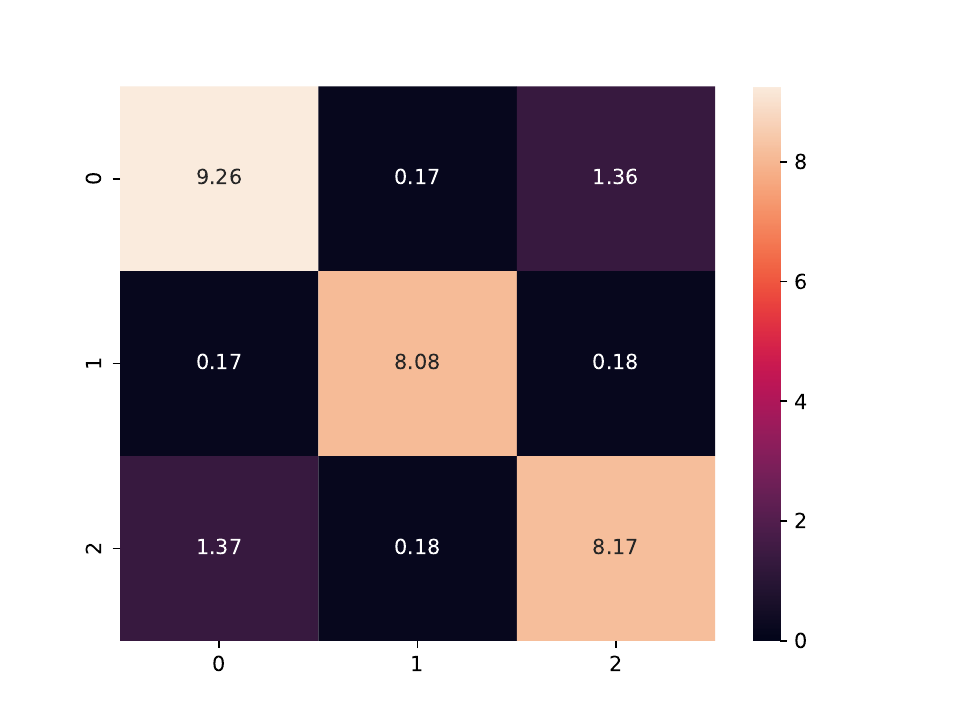}
        \caption{SETransformer Component-wise Performance}
    \end{subfigure}
    \begin{subfigure}{0.45\textwidth}
        \includegraphics[width=\textwidth]{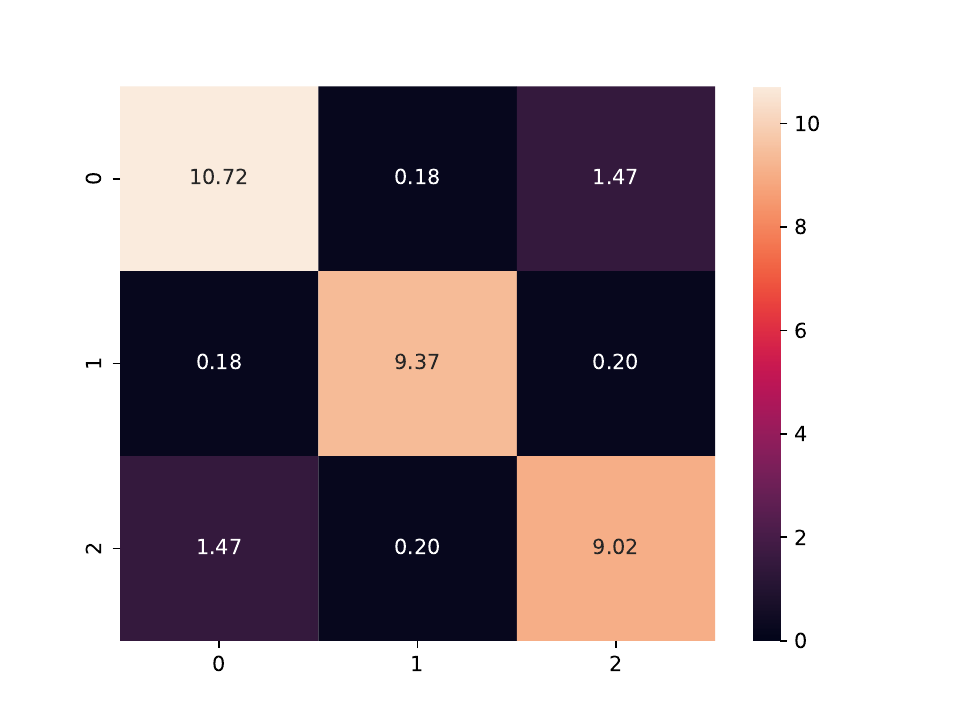}
        \caption{SEGNN Component-wise Performance}
    \end{subfigure}
    \label{BG-DT}
\end{figure}

\newpage
\subsection{Dielectric Tensor Prediction from Formation Energy}

\begin{figure}[h]
    \centering
    \begin{subfigure}{0.45\textwidth}
        \includegraphics[width=\textwidth]{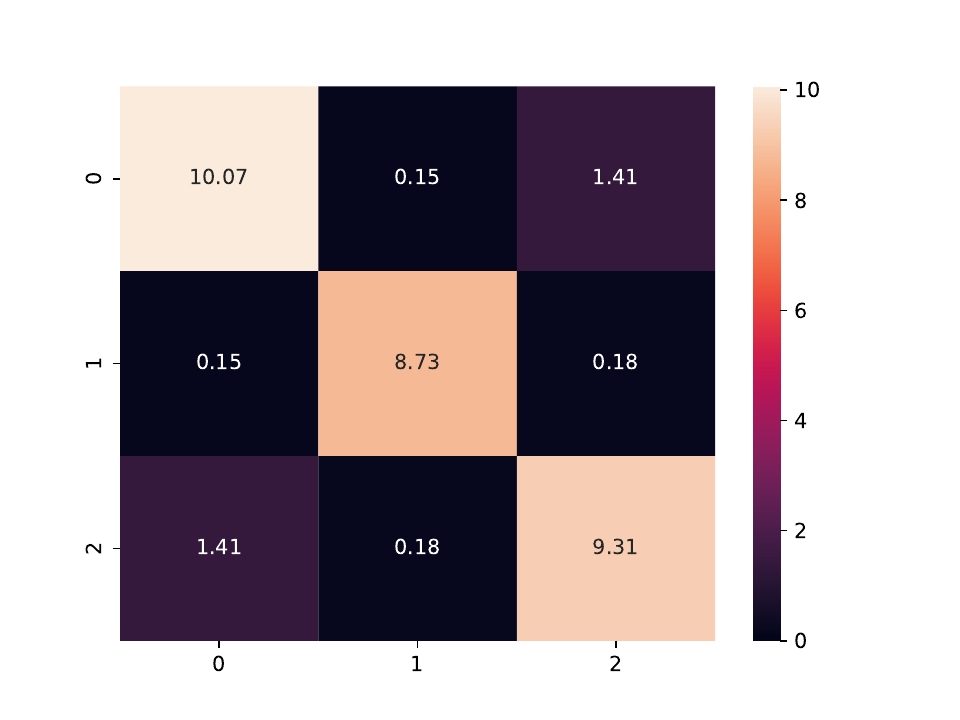}
        \caption{SEConv Component-wise Performance}
    \end{subfigure}
    \begin{subfigure}{0.45\textwidth}
        \includegraphics[width=\textwidth]{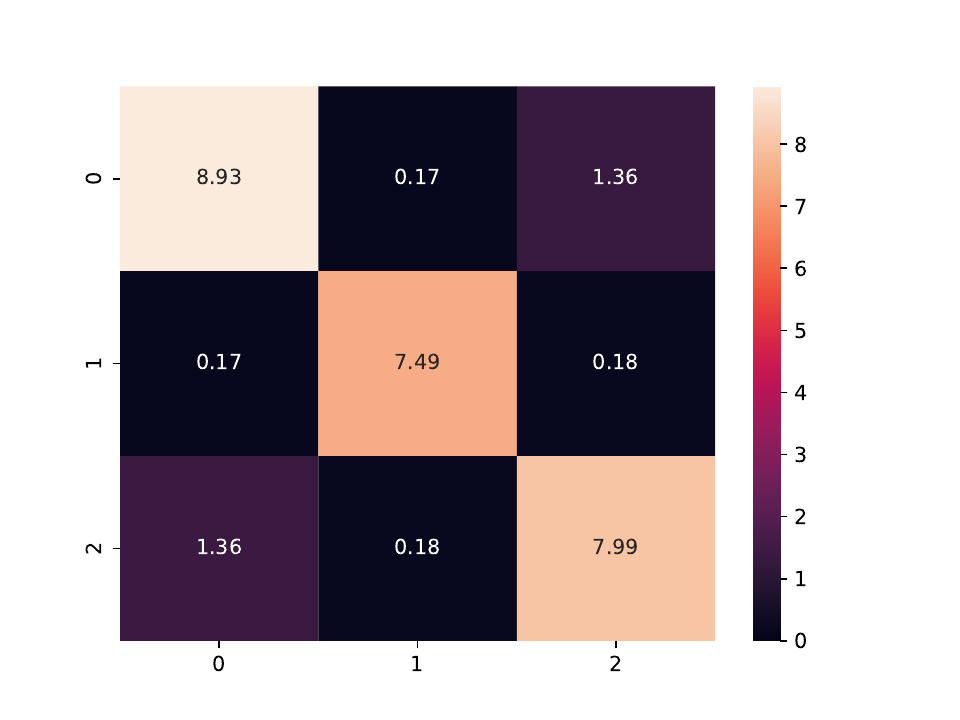}
        \caption{SETransformer Component-wise Performance}
    \end{subfigure}
    \begin{subfigure}{0.45\textwidth}
        \includegraphics[width=\textwidth]{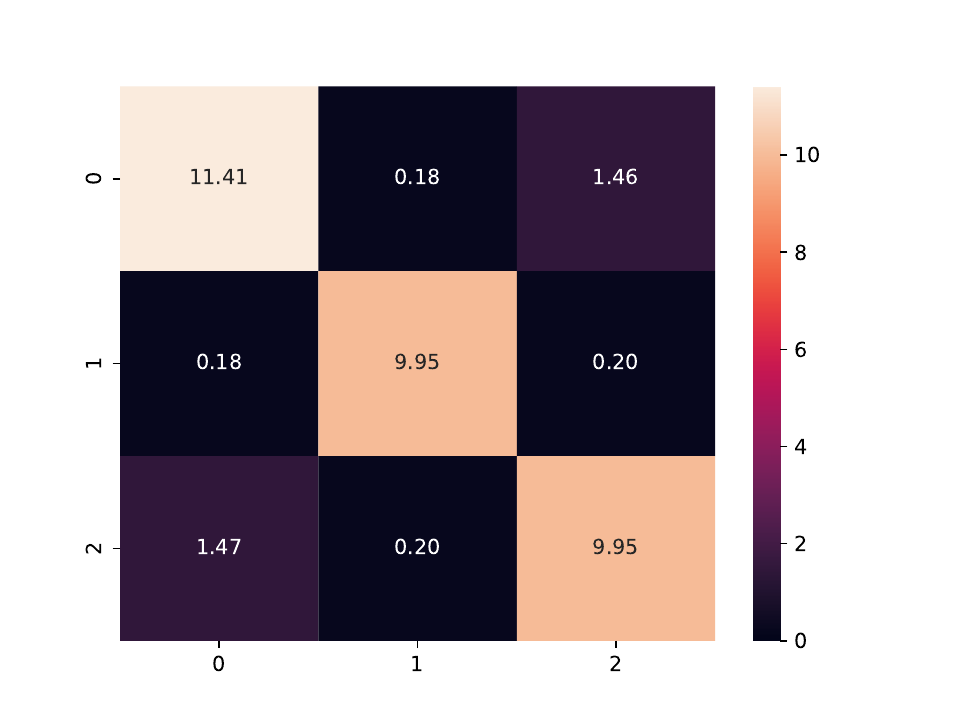}
        \caption{SEGNN Component-wise Performance}
    \end{subfigure}
    \label{FE-DT}
\end{figure}

\newpage\subsection{Dielectric Tensor Prediction From Experiment 1}

\subsubsection{Band Gap as the Scalar Pretraining}

\begin{figure}[h]
    \centering
    \begin{subfigure}{0.45\textwidth}
        \includegraphics[width=\textwidth]{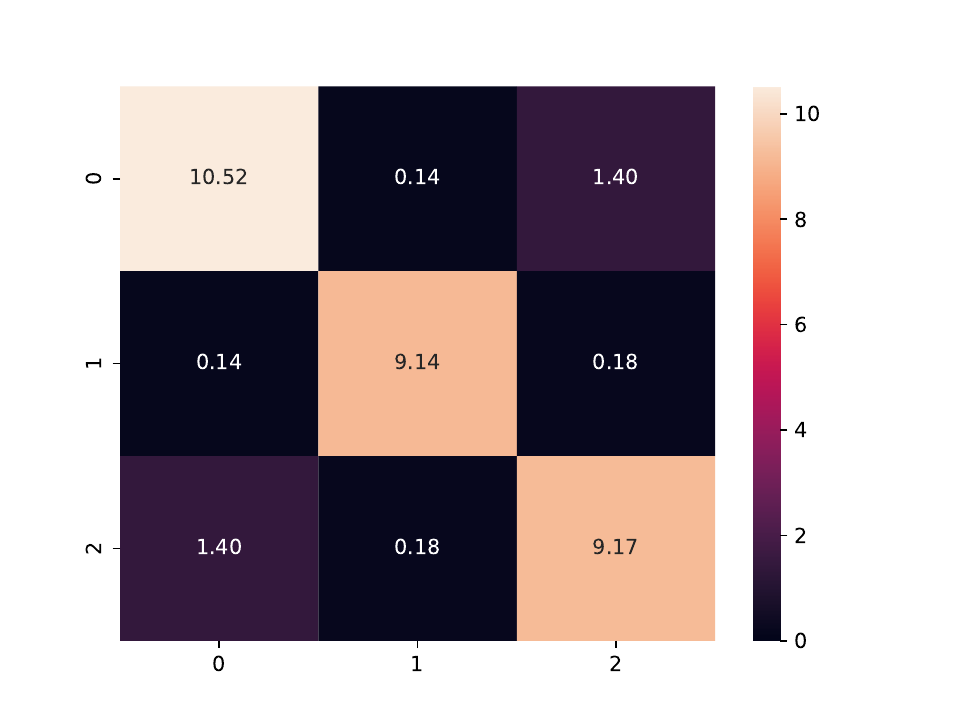}
        \caption{SEConv Component-wise Performance}
    \end{subfigure}
    \begin{subfigure}{0.45\textwidth}
        \includegraphics[width=\textwidth]{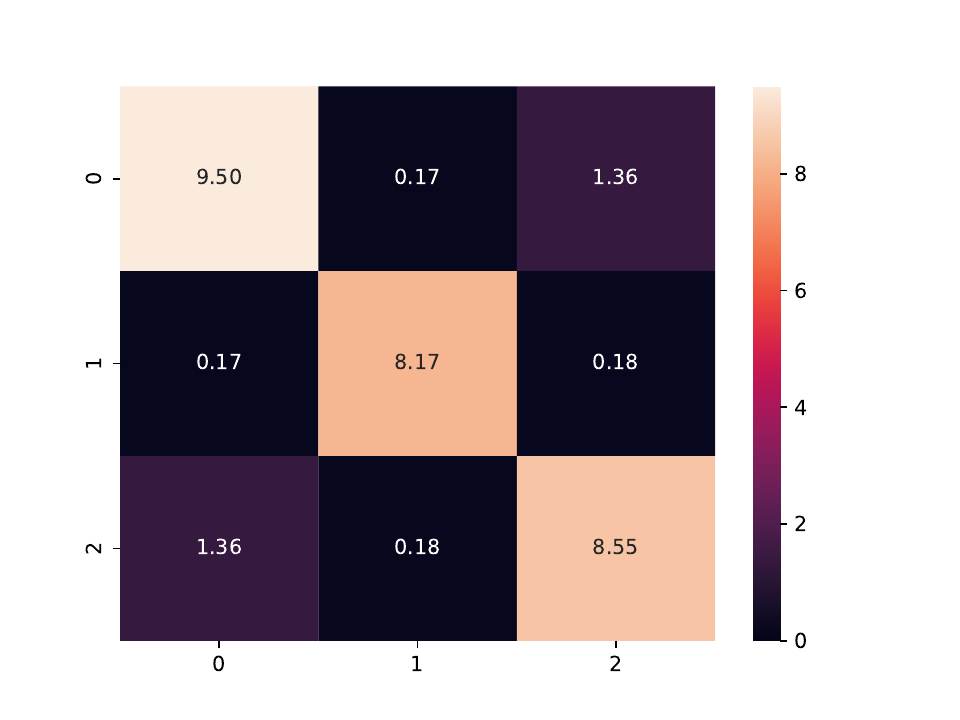}
        \caption{SETransformer Component-wise Performance}
    \end{subfigure}
    \begin{subfigure}{0.45\textwidth}
        \includegraphics[width=\textwidth]{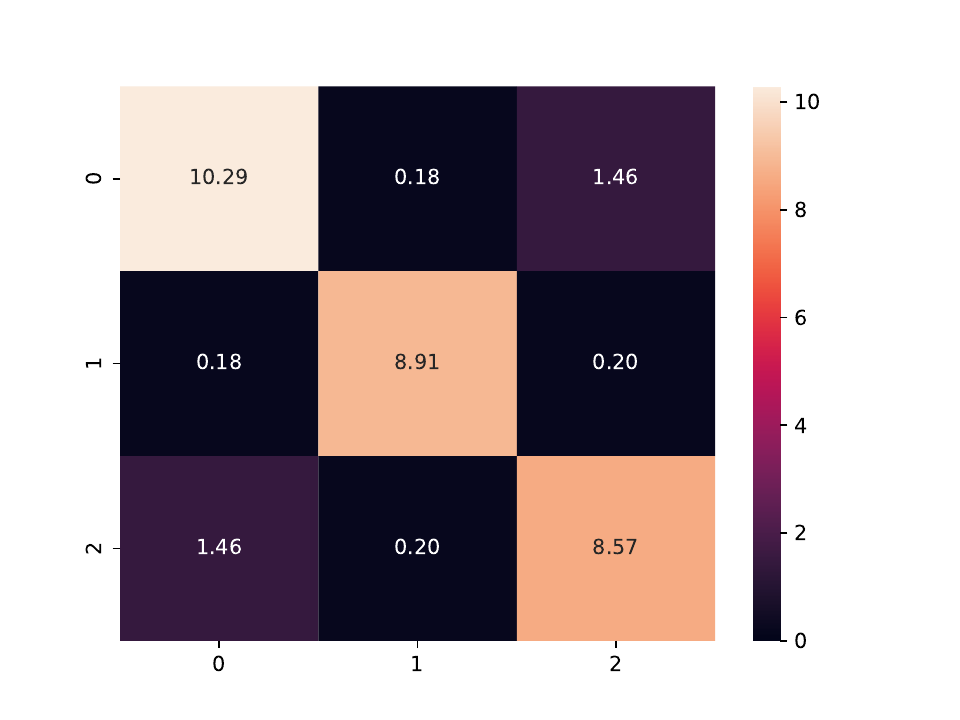}
        \caption{SEGNN Component-wise Performance}
    \end{subfigure}
    \label{exp1-dt}
\end{figure}

\newpage
\subsubsection{Formation Energy as the Scalar Pretraining}

\begin{figure}[h]
    \centering
    \begin{subfigure}{0.45\textwidth}
        \includegraphics[width=\textwidth]{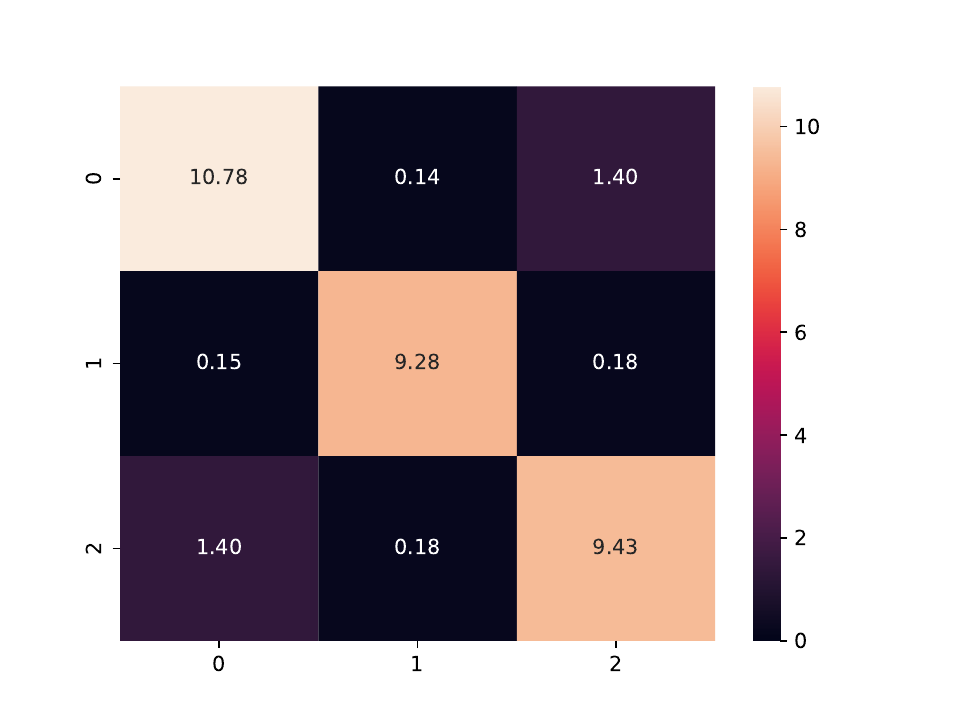}
        \caption{SEConv Component-wise Performance}
    \end{subfigure}
    \begin{subfigure}{0.45\textwidth}
        \includegraphics[width=\textwidth]{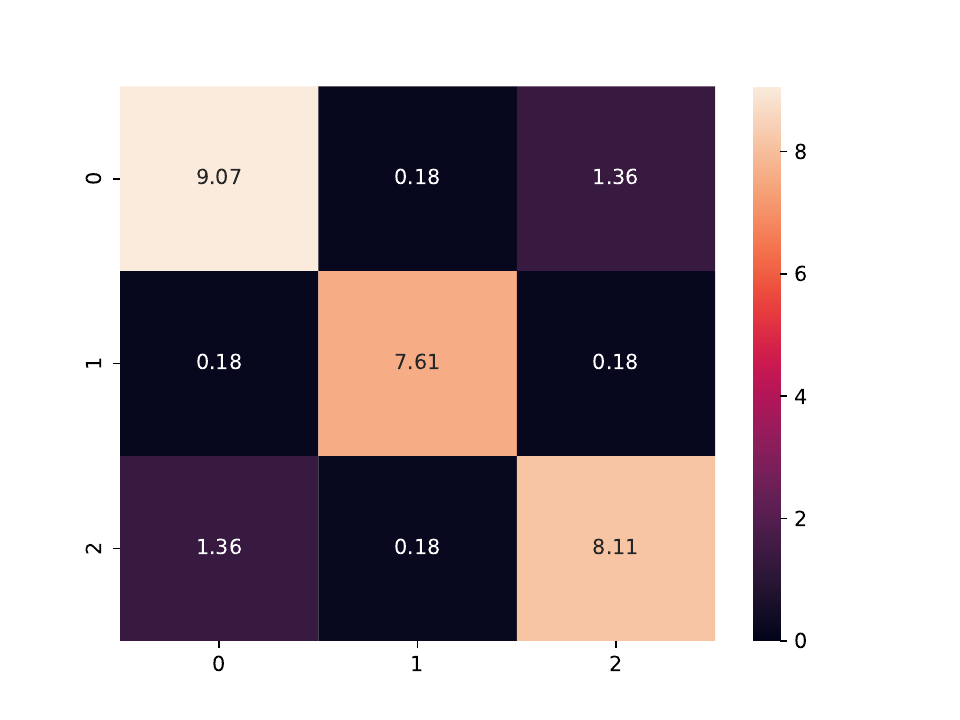}
        \caption{SETransformer Component-wise Performance}
    \end{subfigure}
    \begin{subfigure}{0.45\textwidth}
        \includegraphics[width=\textwidth]{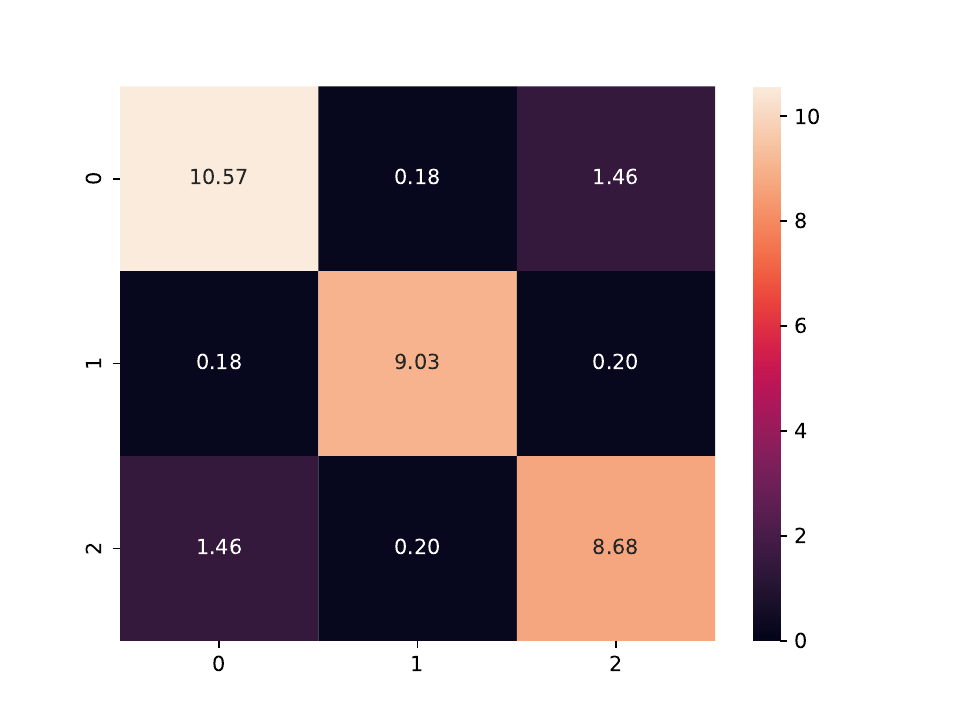}
        \caption{SEGNN Component-wise Performance}
    \end{subfigure}
    \label{exp1-fe-dt}
\end{figure}

\newpage
\section{Piezoelectric Tensor Component-wise MAE}

\subsection{Piezoelectric Tensor Prediction from Band Gap}

\begin{figure}[h]
    \centering
    \begin{subfigure}{0.45\textwidth}
        \includegraphics[width=\textwidth]{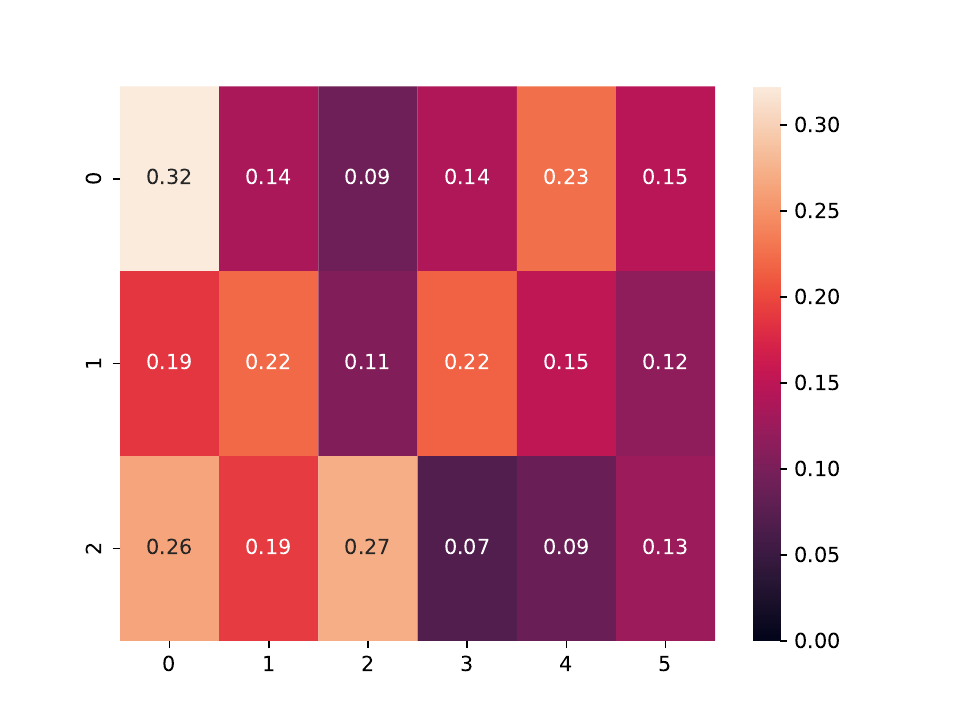}
        \caption{SEConv Component-wise Performance}
    \end{subfigure}
    \begin{subfigure}{0.45\textwidth}
        \includegraphics[width=\textwidth]{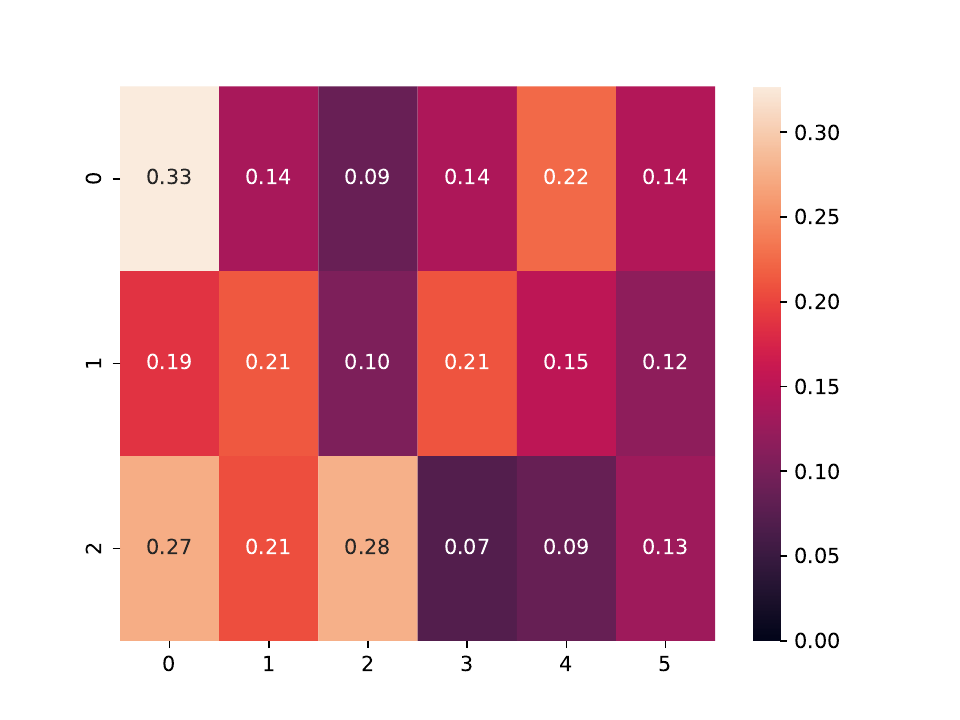}
        \caption{SETransformer Component-wise Performance}
    \end{subfigure}
    \begin{subfigure}{0.45\textwidth}
        \includegraphics[width=\textwidth]{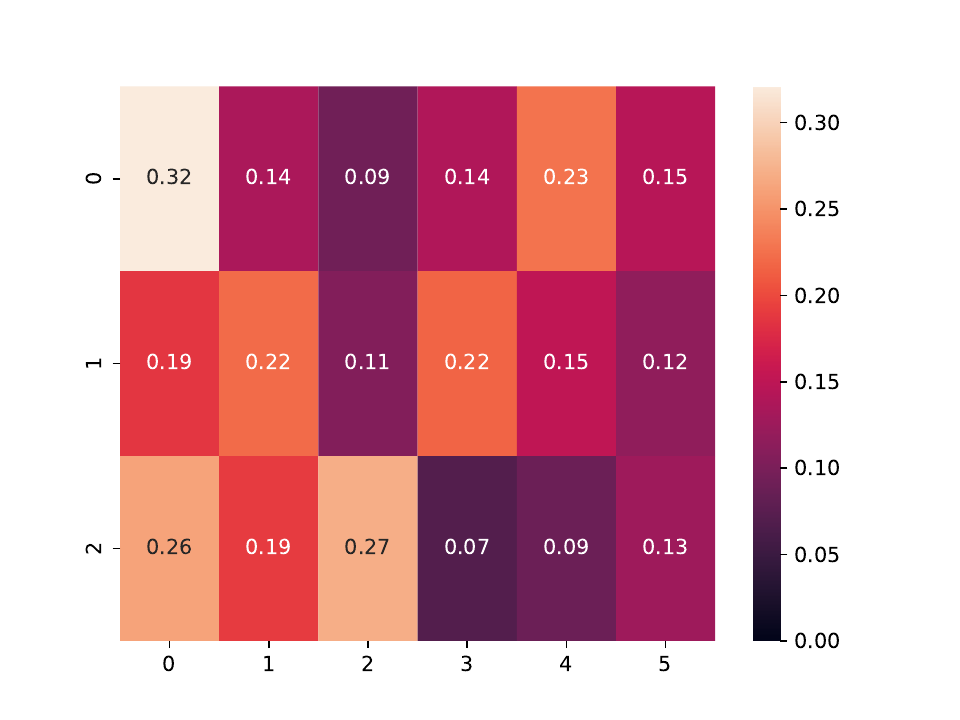}
        \caption{SEGNN Component-wise Performance}
    \end{subfigure}
    \label{BG-PT}
\end{figure}

\newpage
\subsection{Piezoelectric Tensor Prediction from Formation Energy}

\begin{figure}[h]
    \centering
    \begin{subfigure}{0.45\textwidth}
        \includegraphics[width=\textwidth]{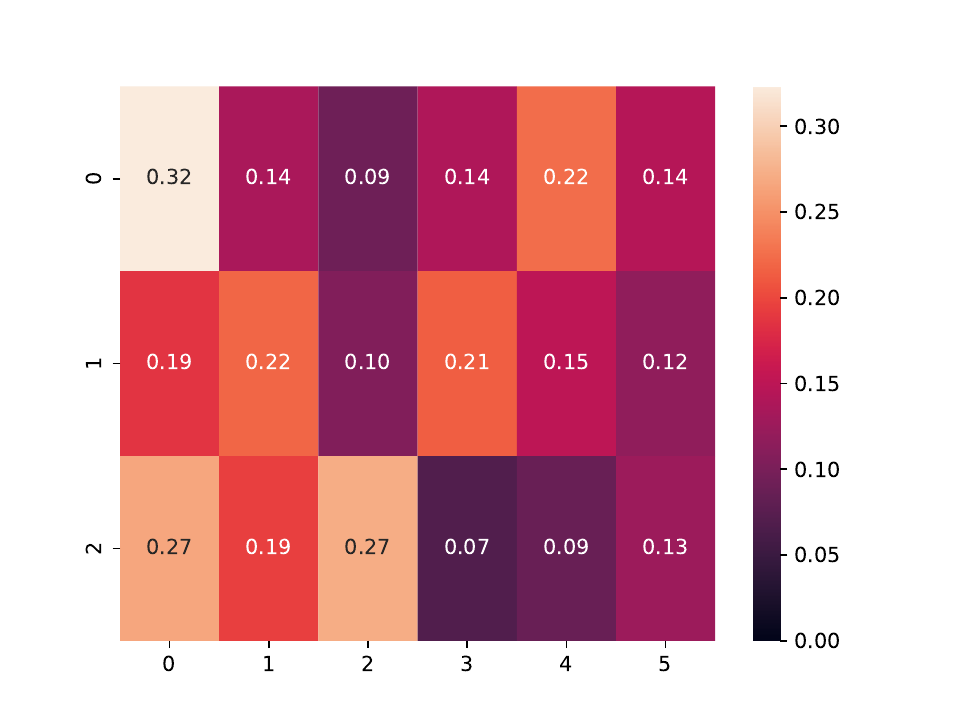}
        \caption{SEConv Component-wise Performance}
    \end{subfigure}
    \begin{subfigure}{0.45\textwidth}
        \includegraphics[width=\textwidth]{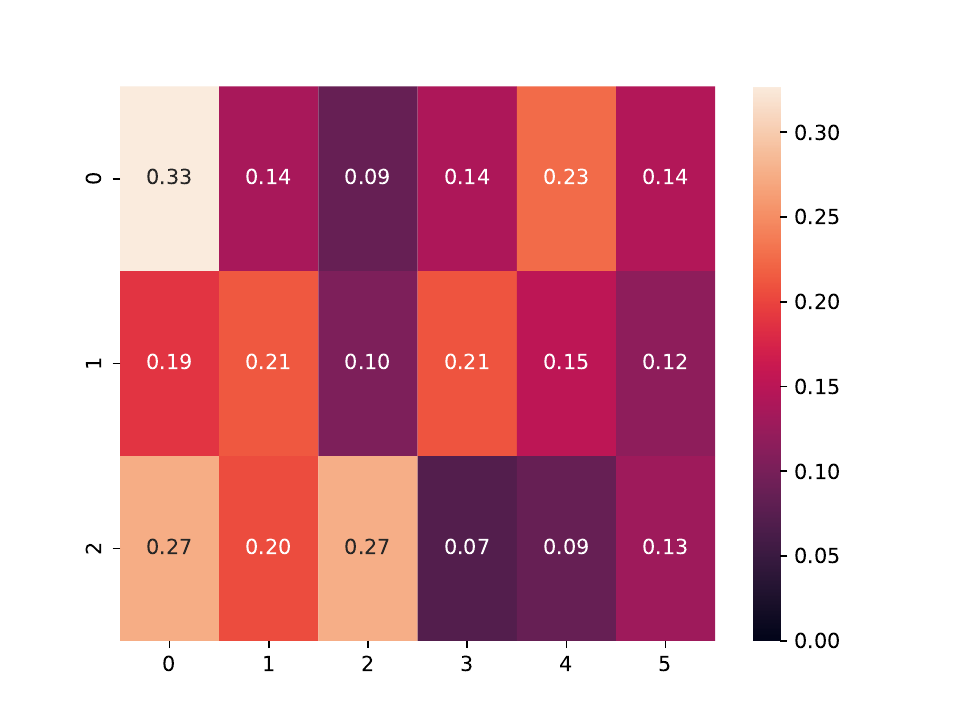}
        \caption{SETransformer Component-wise Performance}
    \end{subfigure}
    \begin{subfigure}{0.45\textwidth}
        \includegraphics[width=\textwidth]{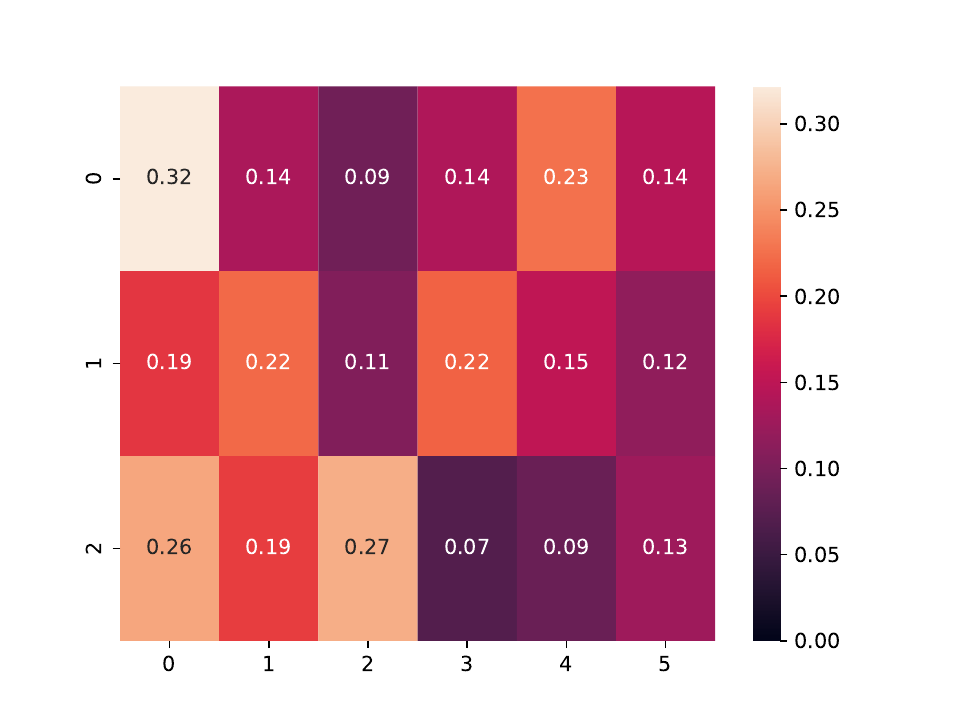}
        \caption{SEGNN Component-wise Performance}
    \end{subfigure}
    \label{FE-PT}
\end{figure}

\newpage
\subsection{Piezoelectric Tensor Prediction From Experiment 1}

\subsubsection{Band Gap as the Scalar Pretraining}

\begin{figure}[h]
    \centering
    \begin{subfigure}{0.45\textwidth}
        \includegraphics[width=\textwidth]{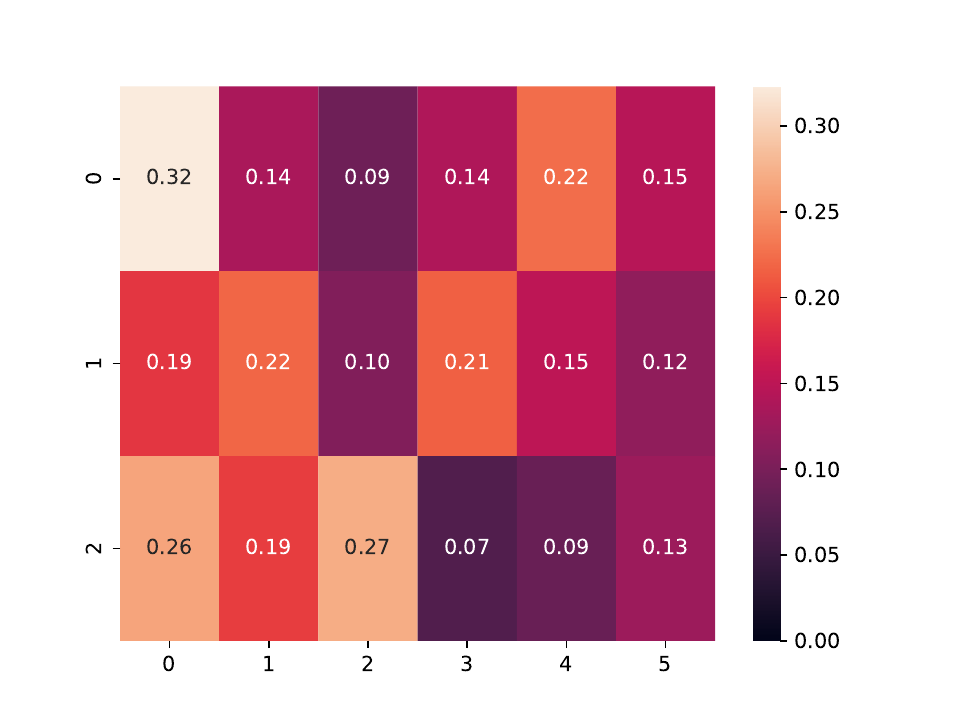}
        \caption{SEConv Component-wise Performance}
    \end{subfigure}
    \begin{subfigure}{0.45\textwidth}
        \includegraphics[width=\textwidth]{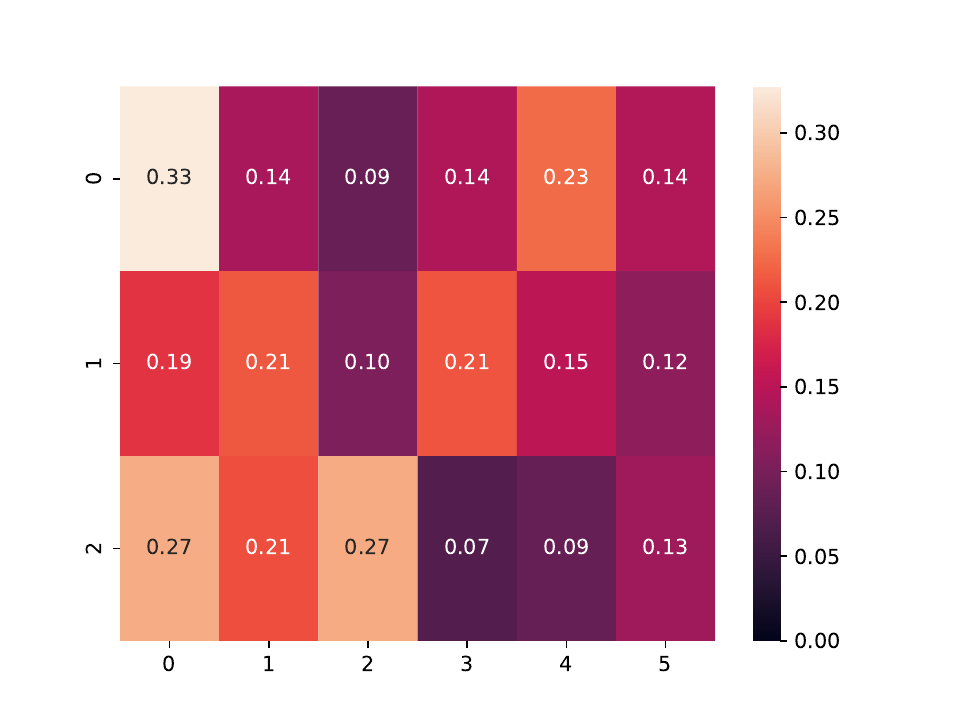}
        \caption{SETransformer Component-wise Performance}
    \end{subfigure}
    \begin{subfigure}{0.45\textwidth}
        \includegraphics[width=\textwidth]{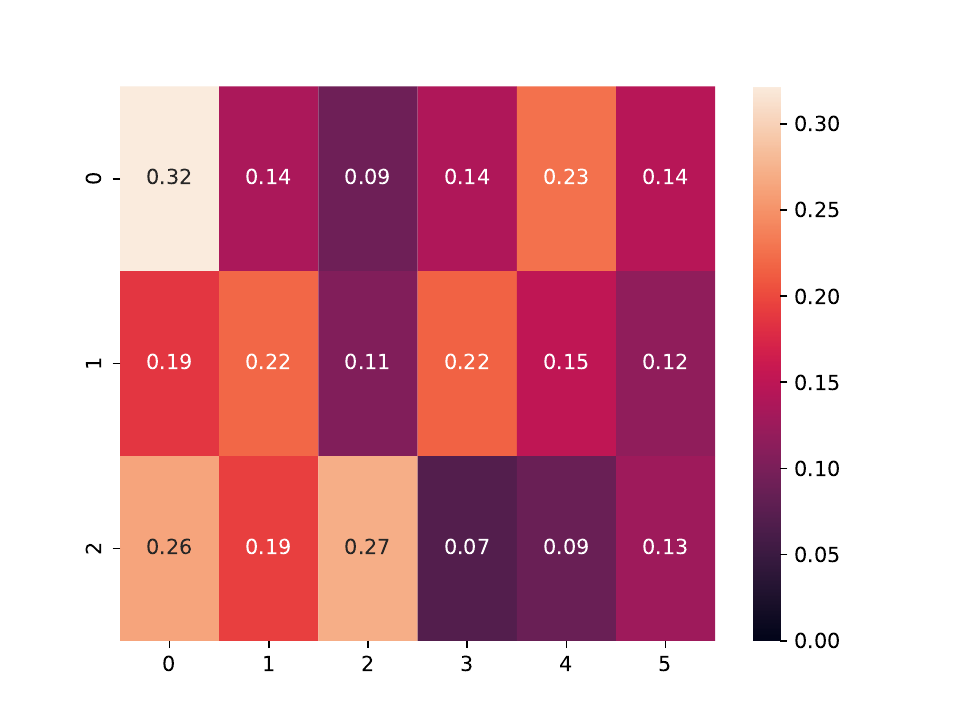}
        \caption{SEGNN Component-wise Performance}
    \end{subfigure}
    \label{exp1-BG-PT}
\end{figure}

\newpage
\subsubsection{Formation Energy as the Scalar Pretraining}

\begin{figure}[h]
    \centering
    \begin{subfigure}{0.45\textwidth}
        \includegraphics[width=\textwidth]{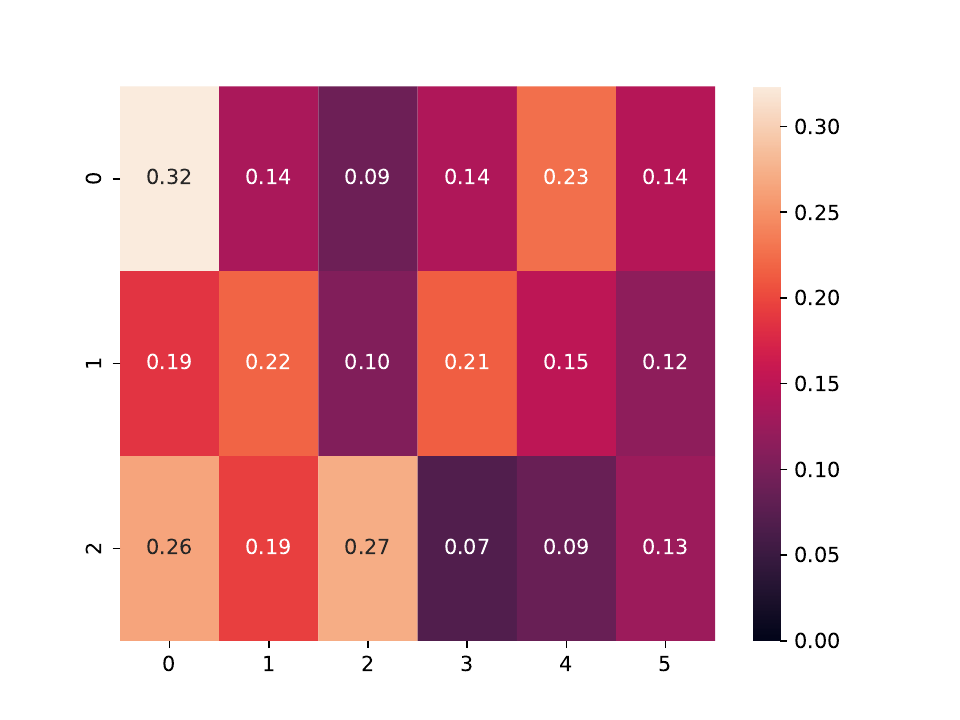}
        \caption{SEConv Component-wise Performance}
    \end{subfigure}
    \begin{subfigure}{0.45\textwidth}
        \includegraphics[width=\textwidth]{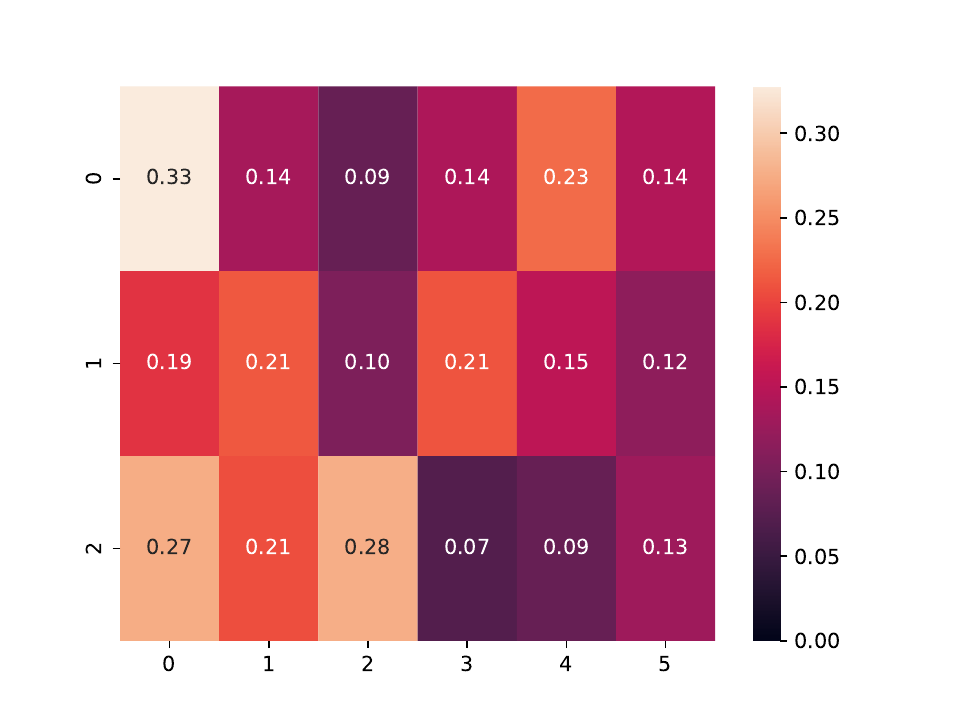}
        \caption{SETransformer Component-wise Performance}
    \end{subfigure}
    \begin{subfigure}{0.45\textwidth}
        \includegraphics[width=\textwidth]{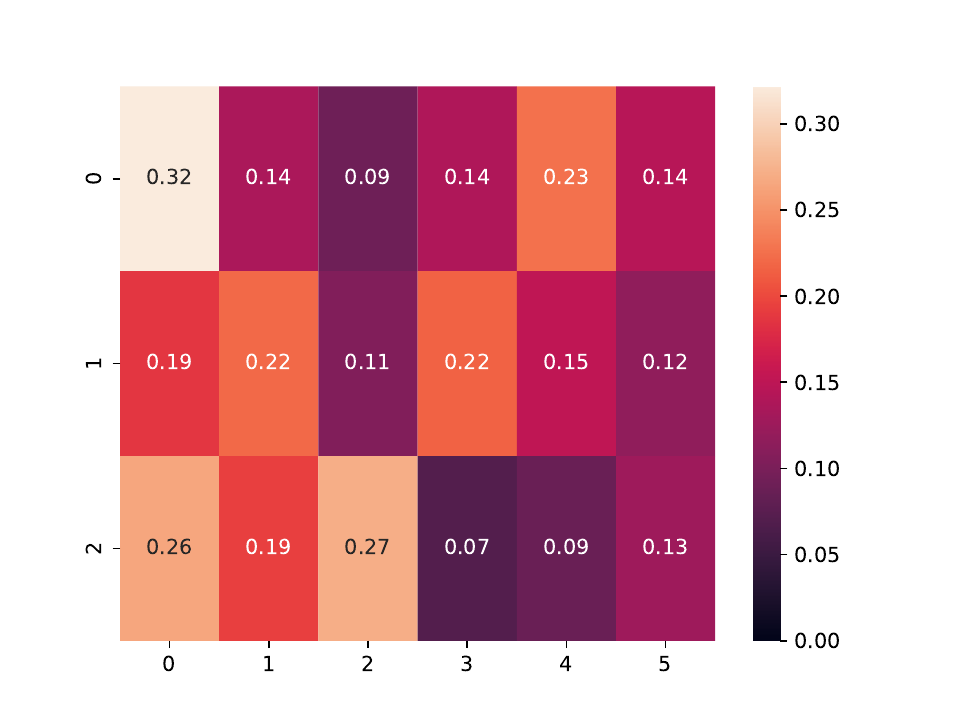}
        \caption{SEGNN Component-wise Performance}
    \end{subfigure}
    \label{exp1-FE-PT}
\end{figure}

\newpage
\subsection{Piezoelectric Tensor Prediction From Experiment 2}

\subsubsection{Band Gap as the Scalar Pretraining}

\begin{figure}[h]
    \centering
    \begin{subfigure}{0.45\textwidth}
        \includegraphics[width=\textwidth]{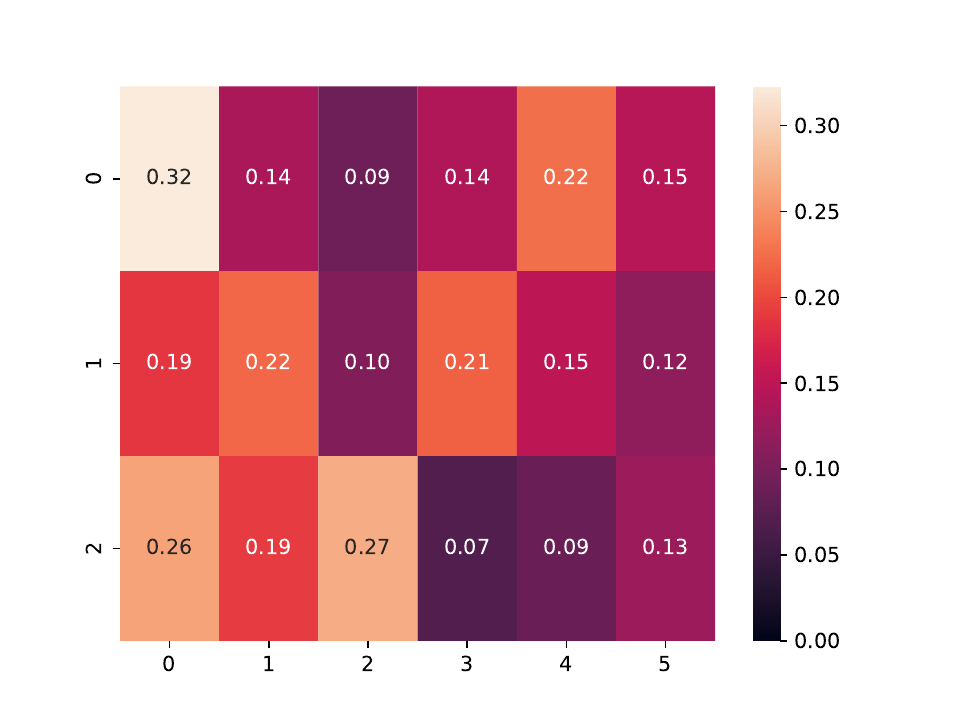}
        \caption{SEConv Component-wise Performance}
    \end{subfigure}
    \begin{subfigure}{0.45\textwidth}
        \includegraphics[width=\textwidth]{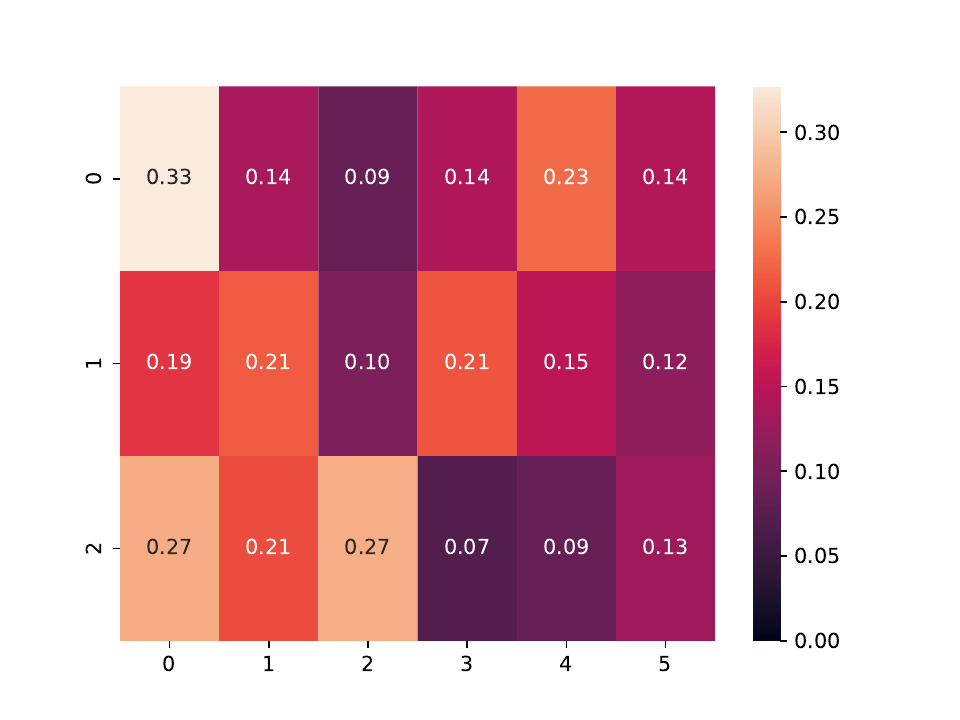}
        \caption{SETransformer Component-wise Performance}
    \end{subfigure}
    \begin{subfigure}{0.45\textwidth}
        \includegraphics[width=\textwidth]{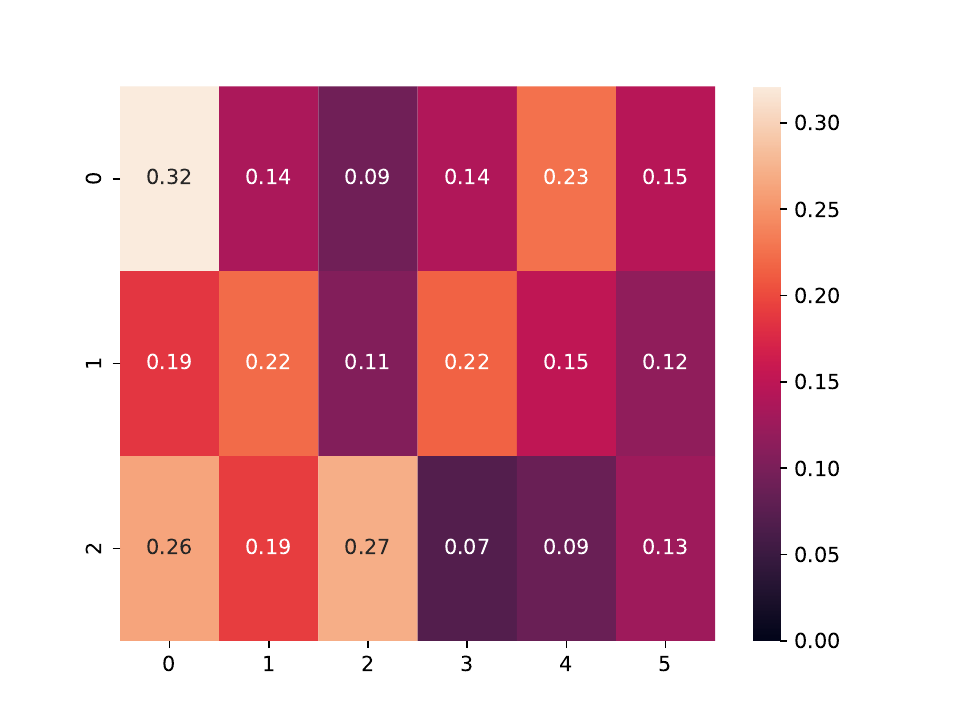}
        \caption{SEGNN Component-wise Performance}
    \end{subfigure}
    \label{exp2-BG-PT}
\end{figure}

\newpage
\subsubsection{Formation Energy as the Scalar Pretraining}

\begin{figure}[h]
    \centering
    \begin{subfigure}{0.45\textwidth}
        \includegraphics[width=\textwidth]{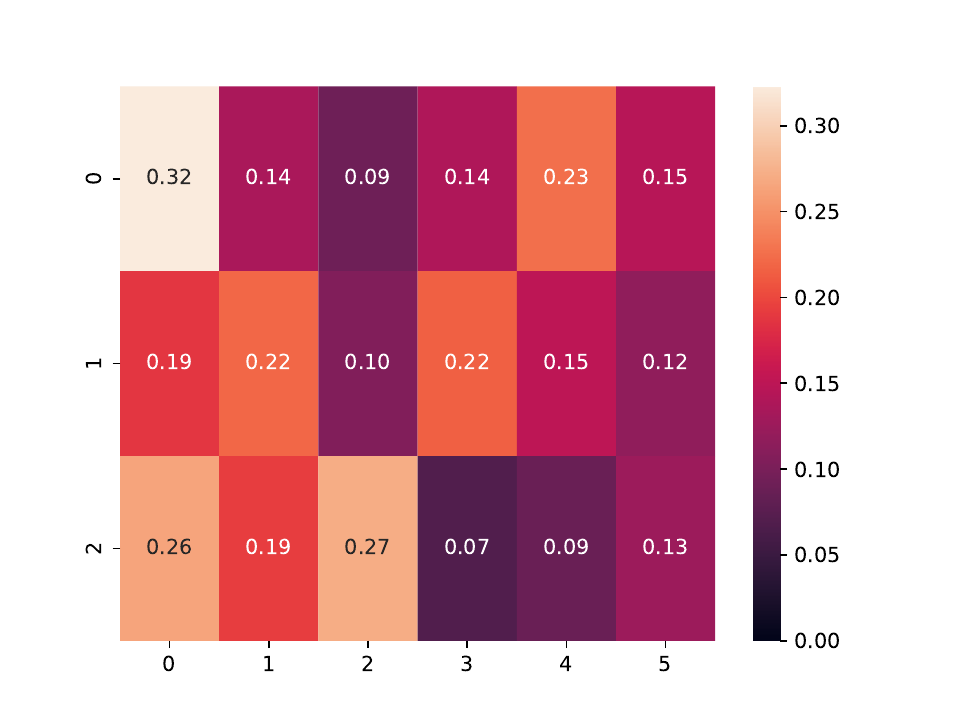}
        \caption{SEConv Component-wise Performance}
    \end{subfigure}
    \begin{subfigure}{0.45\textwidth}
        \includegraphics[width=\textwidth]{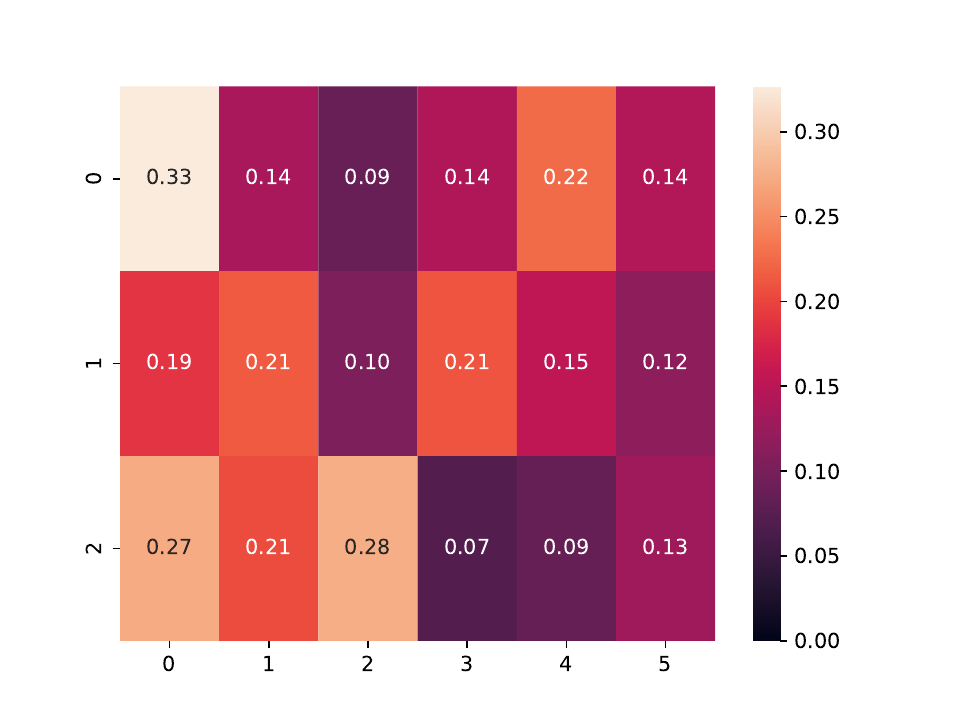}
        \caption{SETransformer Component-wise Performance}
    \end{subfigure}
    \begin{subfigure}{0.45\textwidth}
        \includegraphics[width=\textwidth]{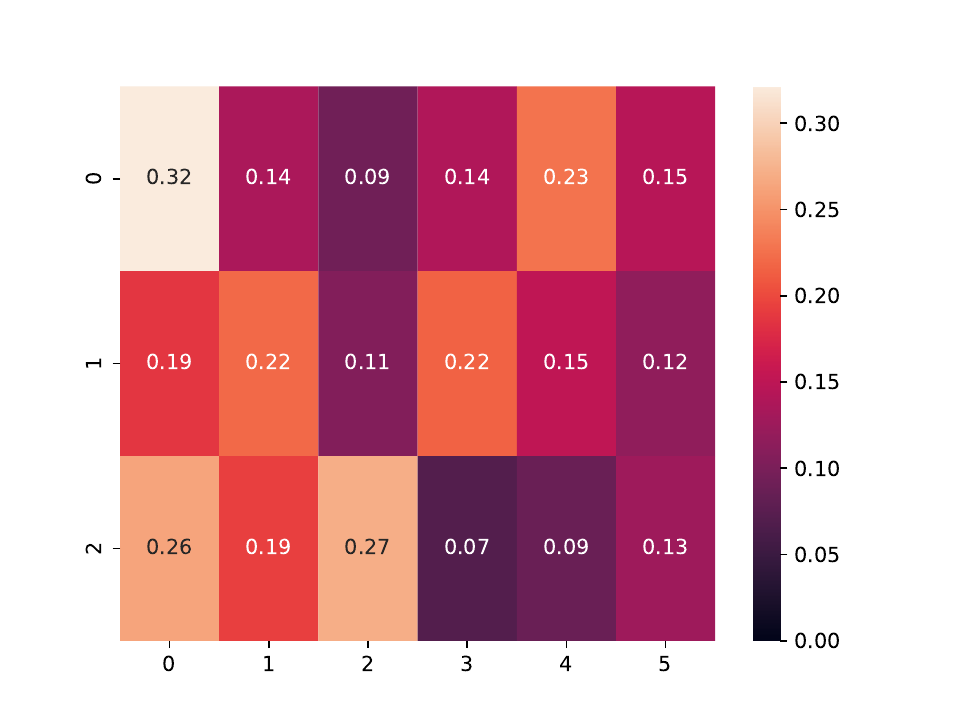}
        \caption{SEGNN Component-wise Performance}
    \end{subfigure}
    \label{exp2-FE-PT}
\end{figure}

\newpage
\subsection{Piezoelectric Tensor Prediction From Experiment 3}

\subsubsection{Band Gap as the Scalar Pretraining}

\begin{figure}[h]
    \centering
    \begin{subfigure}{0.45\textwidth}
        \includegraphics[width=\textwidth]{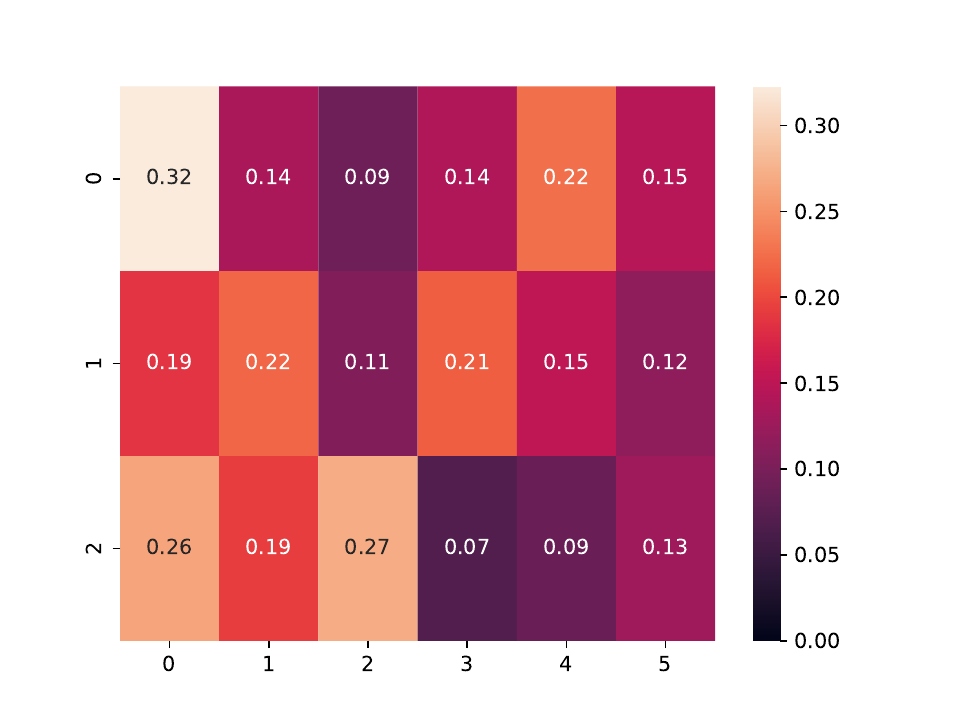}
        \caption{SEConv Component-wise Performance}
    \end{subfigure}
    \begin{subfigure}{0.45\textwidth}
        \includegraphics[width=\textwidth]{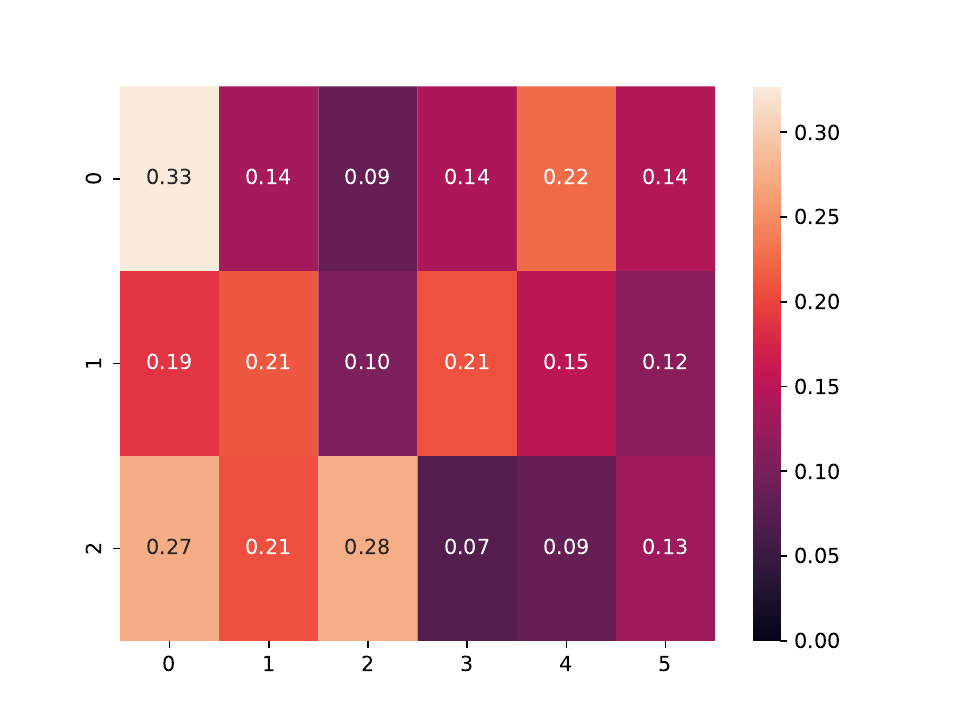}
        \caption{SETransformer Component-wise Performance}
    \end{subfigure}
    \begin{subfigure}{0.45\textwidth}
        \includegraphics[width=\textwidth]{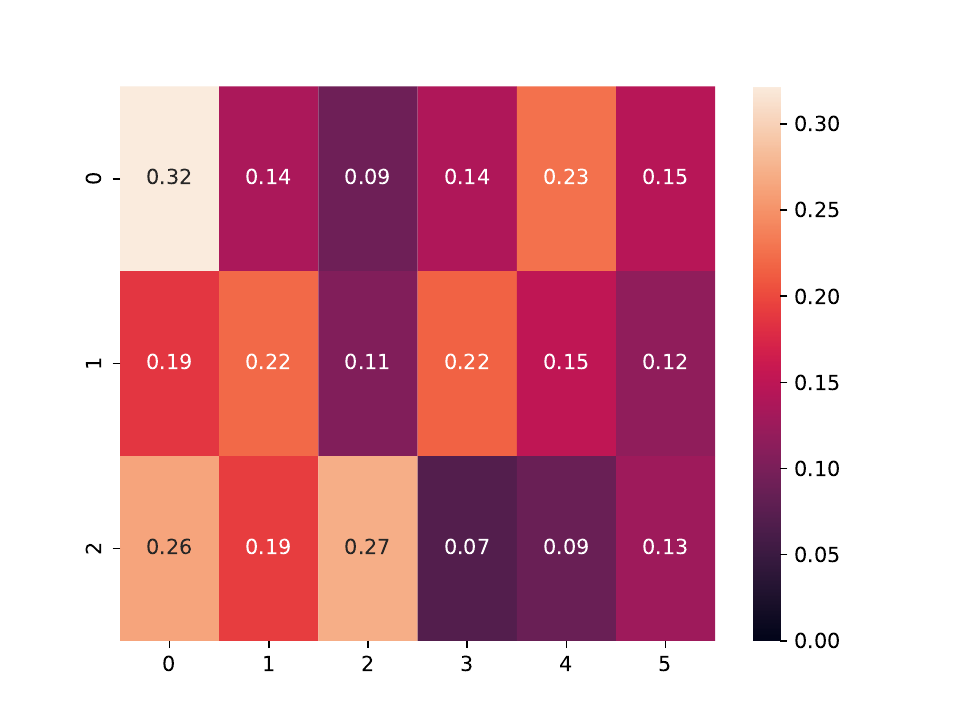}
        \caption{SEGNN Component-wise Performance}
    \end{subfigure}
    \label{exp3-BG-PT}
\end{figure}

\newpage
\subsubsection{Formation Energy as the Scalar Pretraining}

\begin{figure}[h]
    \centering
    \begin{subfigure}{0.45\textwidth}
        \includegraphics[width=\textwidth]{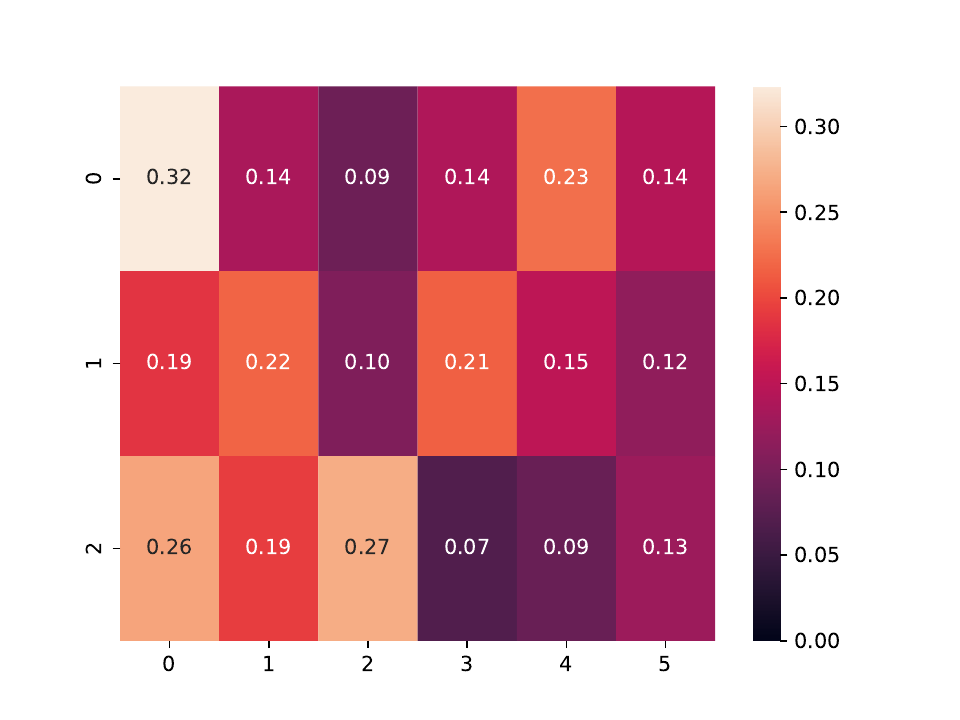}
        \caption{SEConv Component-wise Performance}
    \end{subfigure}
    \begin{subfigure}{0.45\textwidth}
        \includegraphics[width=\textwidth]{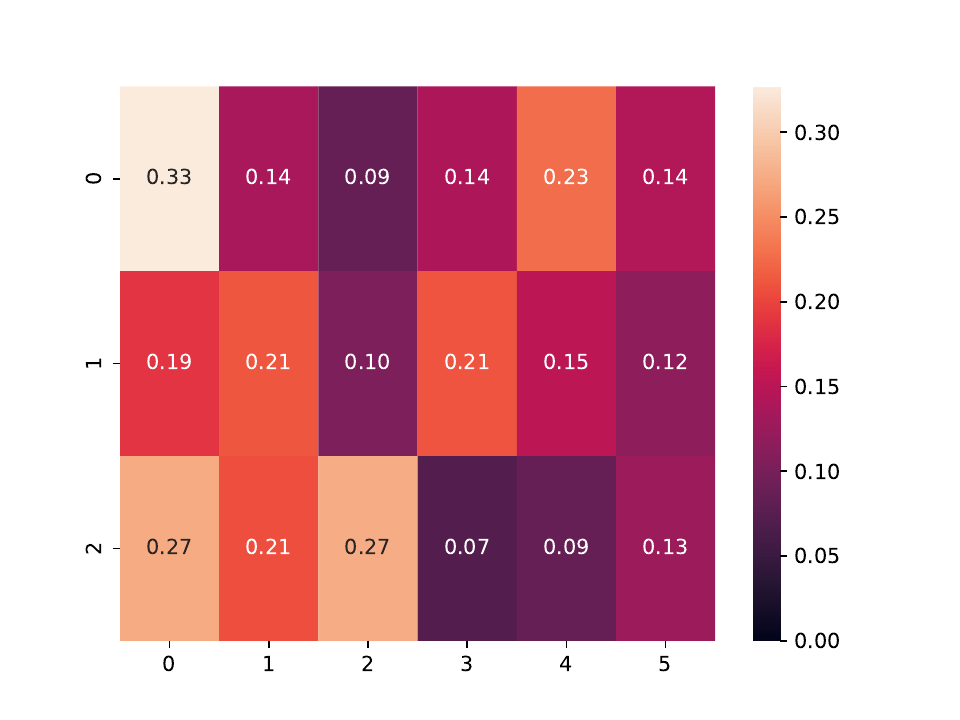}
        \caption{SETransformer Component-wise Performance}
    \end{subfigure}
    \begin{subfigure}{0.45\textwidth}
        \includegraphics[width=\textwidth]{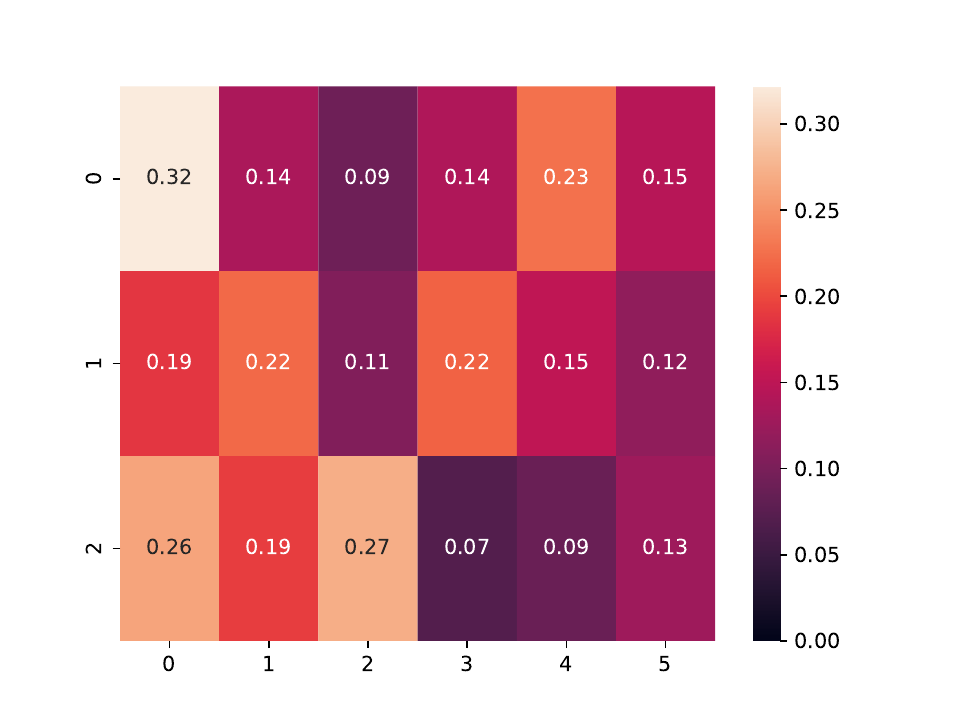}
        \caption{SEGNN Component-wise Performance}
    \end{subfigure}
    \label{exp3-FE-PT}
\end{figure}

\end{document}